\documentclass{amsart}

\usepackage{graphicx}
\usepackage{url}
\usepackage{enumerate}
\usepackage{bm}
\usepackage{color}
\usepackage{palatino}
\usepackage{cite} 
\usepackage{setspace}
\usepackage{multirow}
\usepackage{caption}
\usepackage{subcaption}

\newcommand{\citep}{\cite}
\newcommand{\citet}{\cite}

\DeclareMathOperator{\Poisson}{Poisson}
\DeclareMathOperator{\diag}{diag}

\newcommand{\DGamma}{\mathrm{DGamma}}

\newcommand{\bprim}{\beta'}
\newcommand{\ahat}{\hat{\alpha}}
\newcommand{\bhat}{\hat{\beta}}

\newcommand{\arxiv}[1]{#1}
\newcommand{\notarxiv}[1]{}
\newcommand{\eat}[1]{}

\newcommand{\beginsupplement}{%
        \setcounter{table}{0}
        \renewcommand{\thetable}{S\arabic{table}}%
        \setcounter{figure}{0}
        \renewcommand{\thefigure}{S\arabic{figure}}%
     }

\newcommand{\FIGoverview}{\
\begin{figure}
  \arxiv{\includegraphics[width=0.8 \textwidth]{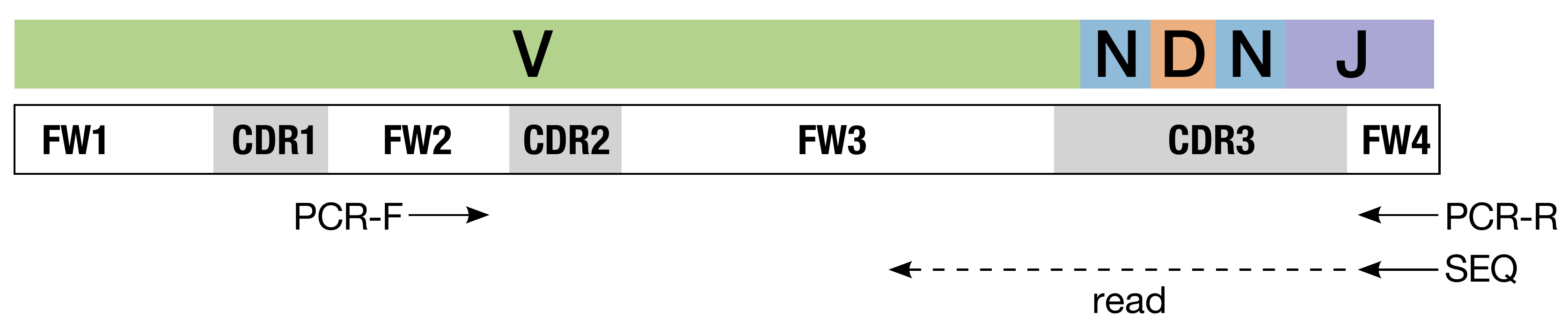}}
\caption{\
  B cell receptor schematic showing variable (V), diversity (D) and joining (J) gene segments as well as framework (FW) and complementarity-determining regions (CDRs).
  In VDJ recombination, individual V, D, and J gene segments are randomly selected are joined together via a process that deletes some randomly distributed number of nucleotides on their boundaries then joins them together with random ``non templated insertions'' (N).
  The specificity of an antibody is primarily determined by the region defined by the heavy chain recombination site, referred to as the third complementarity determining region (CDR3).
  The sequence data for this study started in the fourth framework (FW) region and continued into the third.
  Amplification was via a forward primer in the FW2 region and a reverse primer in the FW4 region.
}
\label{FIGoverview}
\end{figure}
}

\newcommand{\FIGcompPca}{\
\begin{figure}
\begin{center}
  \arxiv{\includegraphics[width=0.95 \textwidth]{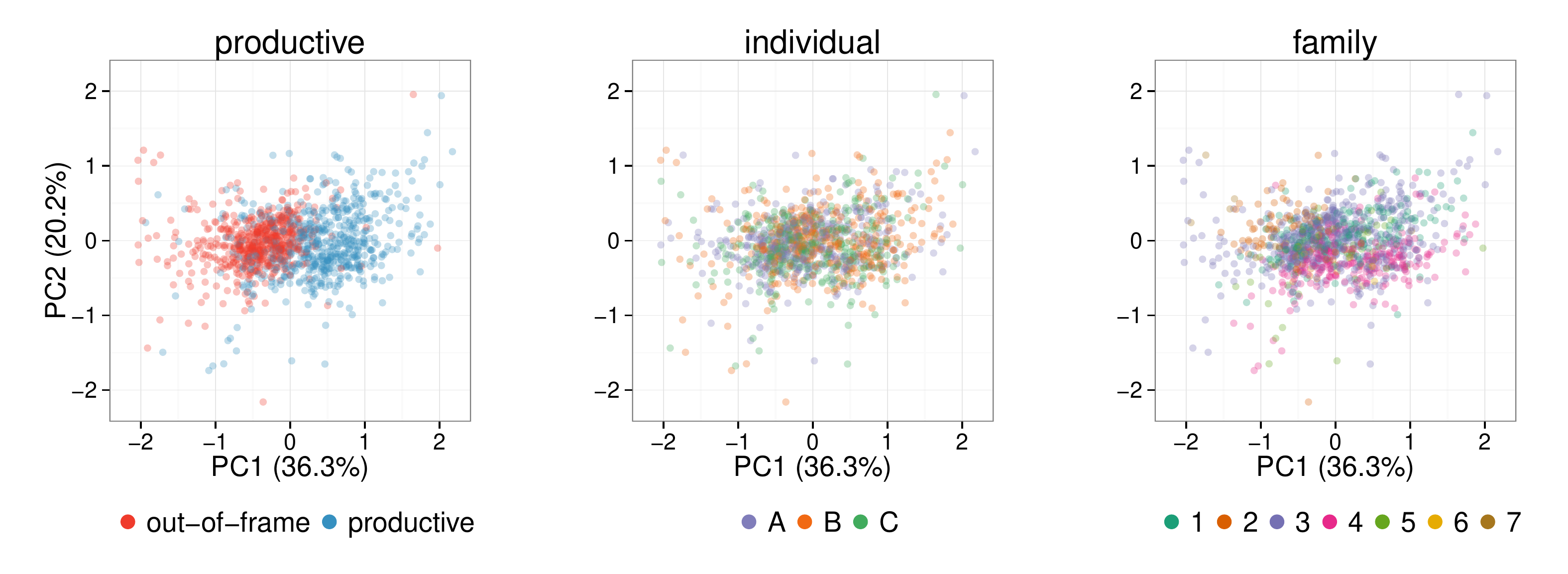}}
\end{center}
\caption{\
  First (x-axis) and second (y-axis) principal components from PCA performed on centered log-transformed median-time transition matrices for V gene segments.
  Points plotted in a random order, with 22 outliers removed for clarity.
}
\label{FIGcompPca}
\end{figure}
}

\newcommand{\FIGmodelVsHammingDist}{\
\begin{figure}
\begin{center}
  \arxiv{\includegraphics[width=5in]{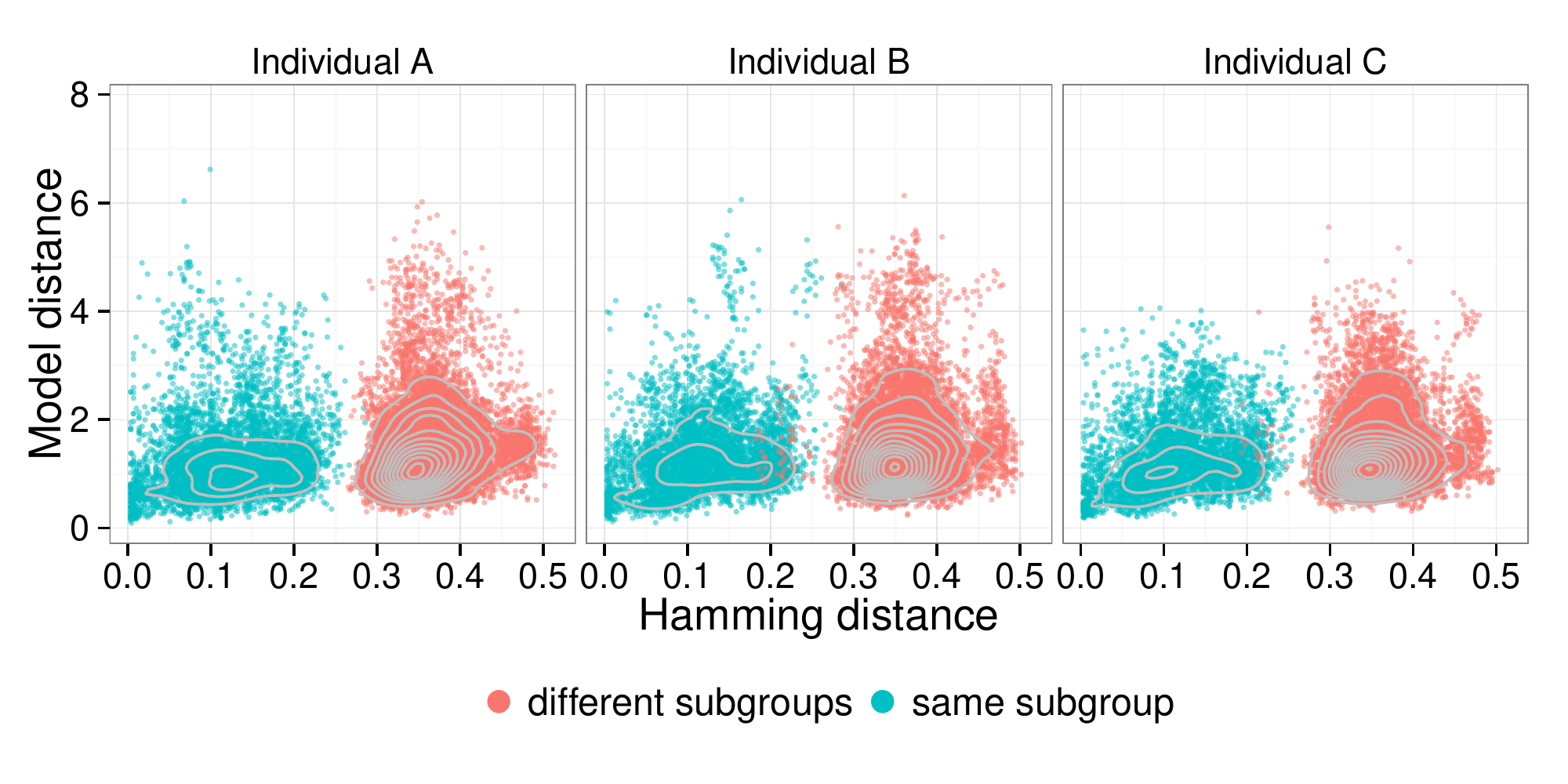}}
\end{center}
\caption{\
  Comparison of Hamming distance between V genes (x-axis) and Euclidean distance between centered log-transformed median time transition matrices for productive rearrangements (y-axis).
  Colors indicate whether the V genes in a comparison come from the same or different subgroups.
  The correlation between the two was significant ($p < 10^{-15}$, Spearman's $\rho = 0.197$).
}
\label{FIGmodelVsHammingDist}
\end{figure}
}

\newcommand{\FIGgtr}{\
\begin{figure}
\begin{center}
  \arxiv{\includegraphics[width=1.0 \textwidth]{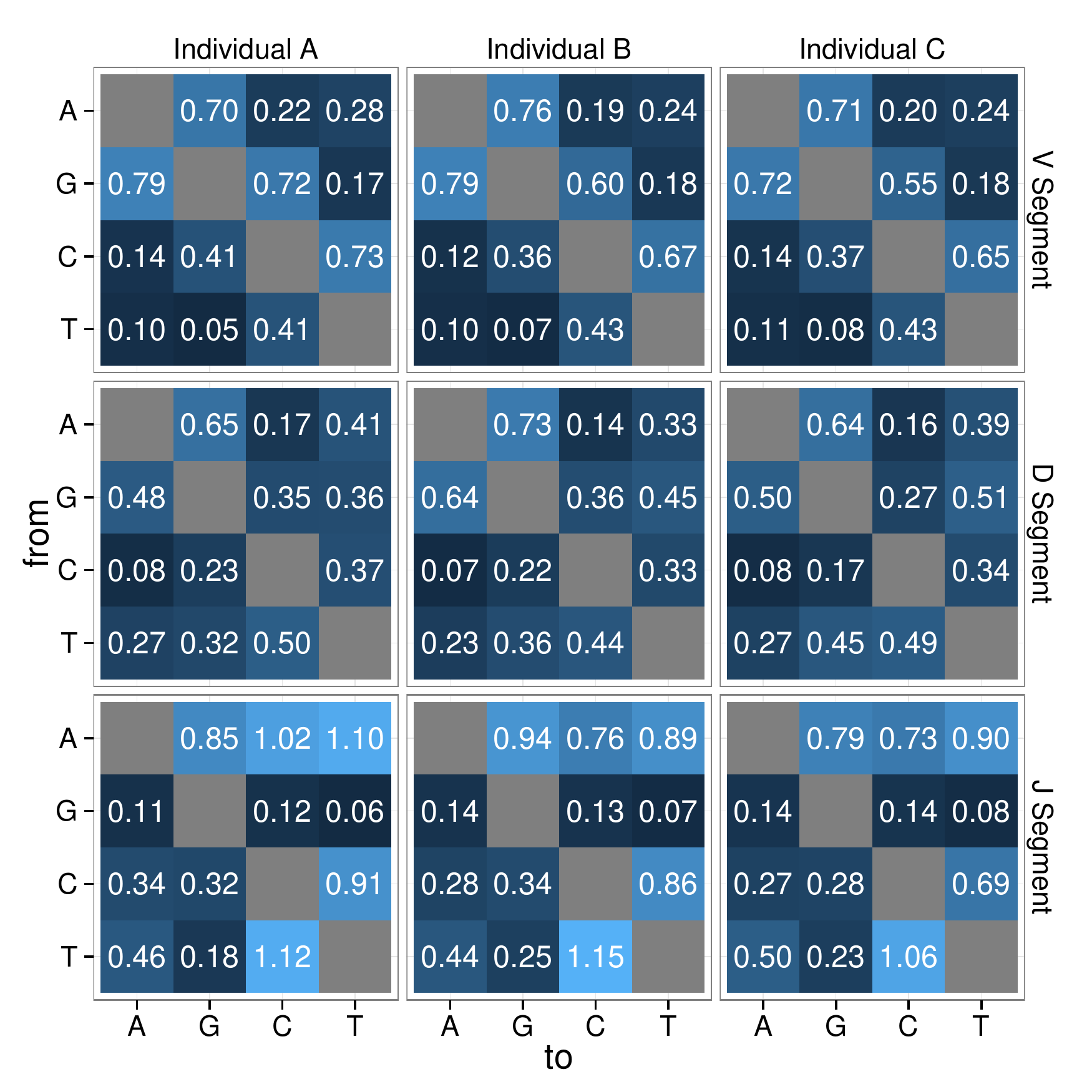}}
\end{center}
\caption{\
  GTR coefficients for the \trqigi\ model estimated under maximum likelihood.
  Rows index the nucleotide found in the germline sequence, whereas columns index the nucleotide found in the observed sequence.
}
\label{FIGgtr}
\end{figure}
}

\newcommand{\FIGgtrBl}{\
\begin{figure}
\begin{center}
  \arxiv{\includegraphics[width=5in]{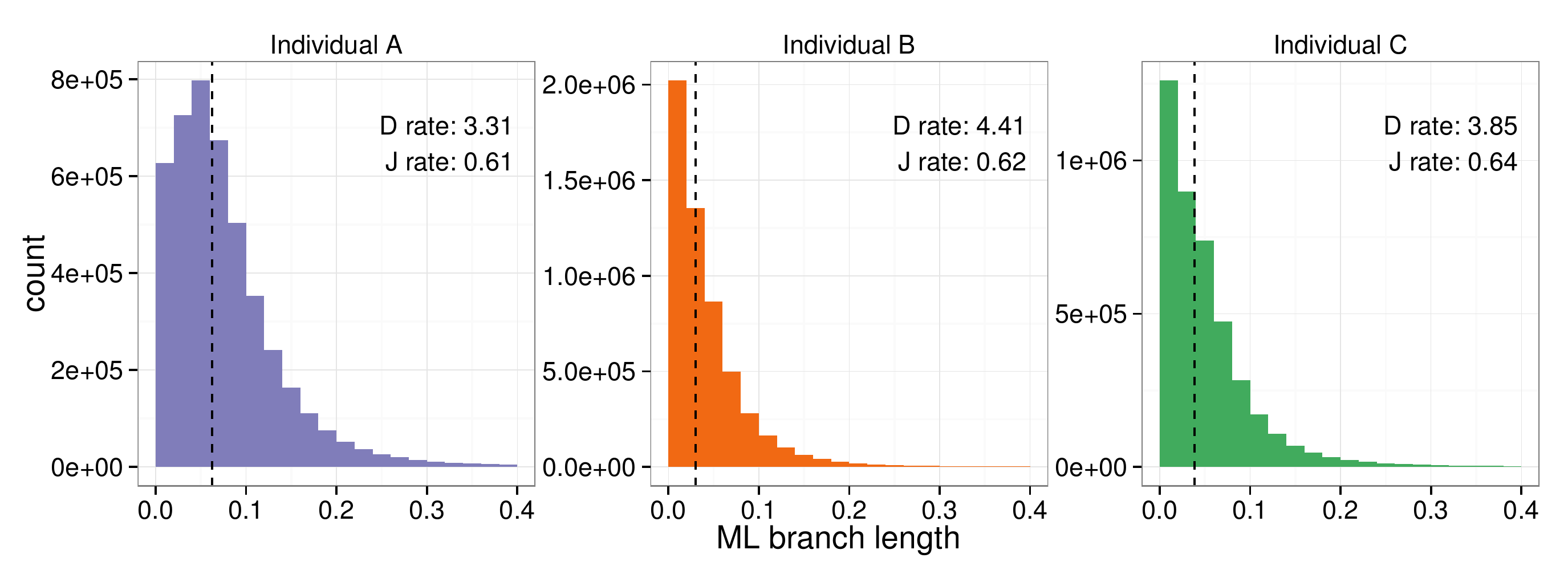}}
\end{center}
\caption{\
  Distribution of maximum likelihood branch lengths estimated under the \trqigi\ model.
  Branch lengths are measured in terms of substitutions per site, and rates given for the D and J segments are relative to a fixed rate of 1 for the V segment.
}
\label{FIGgtrBl}
\end{figure}
}

\newcommand{\FIGdndsNsExample}{\
\begin{figure}
\begin{center}
  \arxiv{\includegraphics[width=1.0 \textwidth]{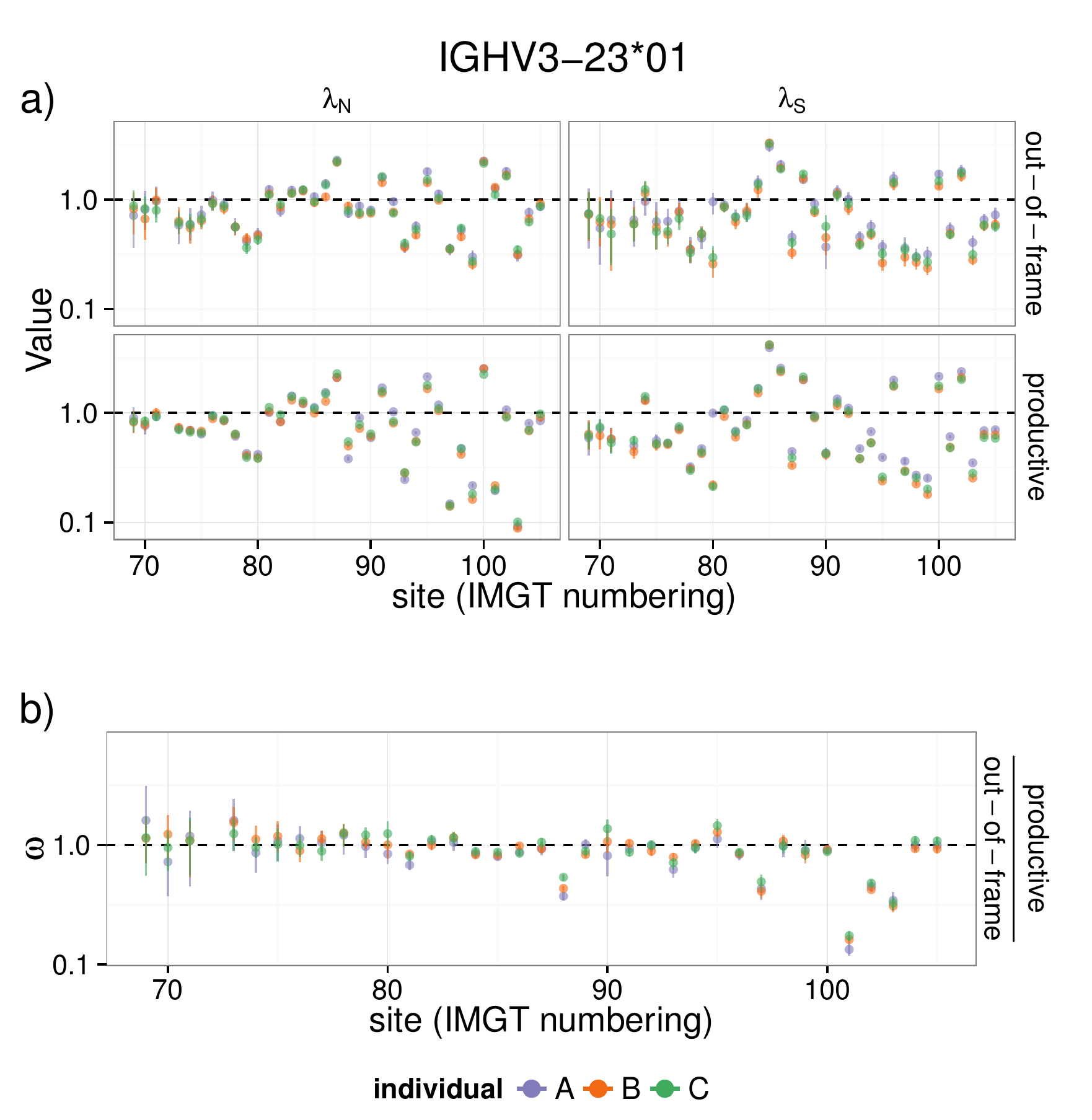}}
\end{center}
\caption{\
  a) Comparison of nonsynonymous ($\lambda_N$) and synonymous ($\lambda_S$) rates in productive and out-of-frame sequences.\
  b) $\omega$ estimates using unproductive rearrangements as a proxy for the neutral process.
  Both panels use data from IGHV3-23*01, the most frequent V gene/allele combination.
}
\label{FIGdndsNsExample}
\end{figure}
}

\newcommand{\FIGdndsDistribution}{\
\begin{figure}ht]
\begin{center}
  \arxiv{\includegraphics[width=1.0 \textwidth]{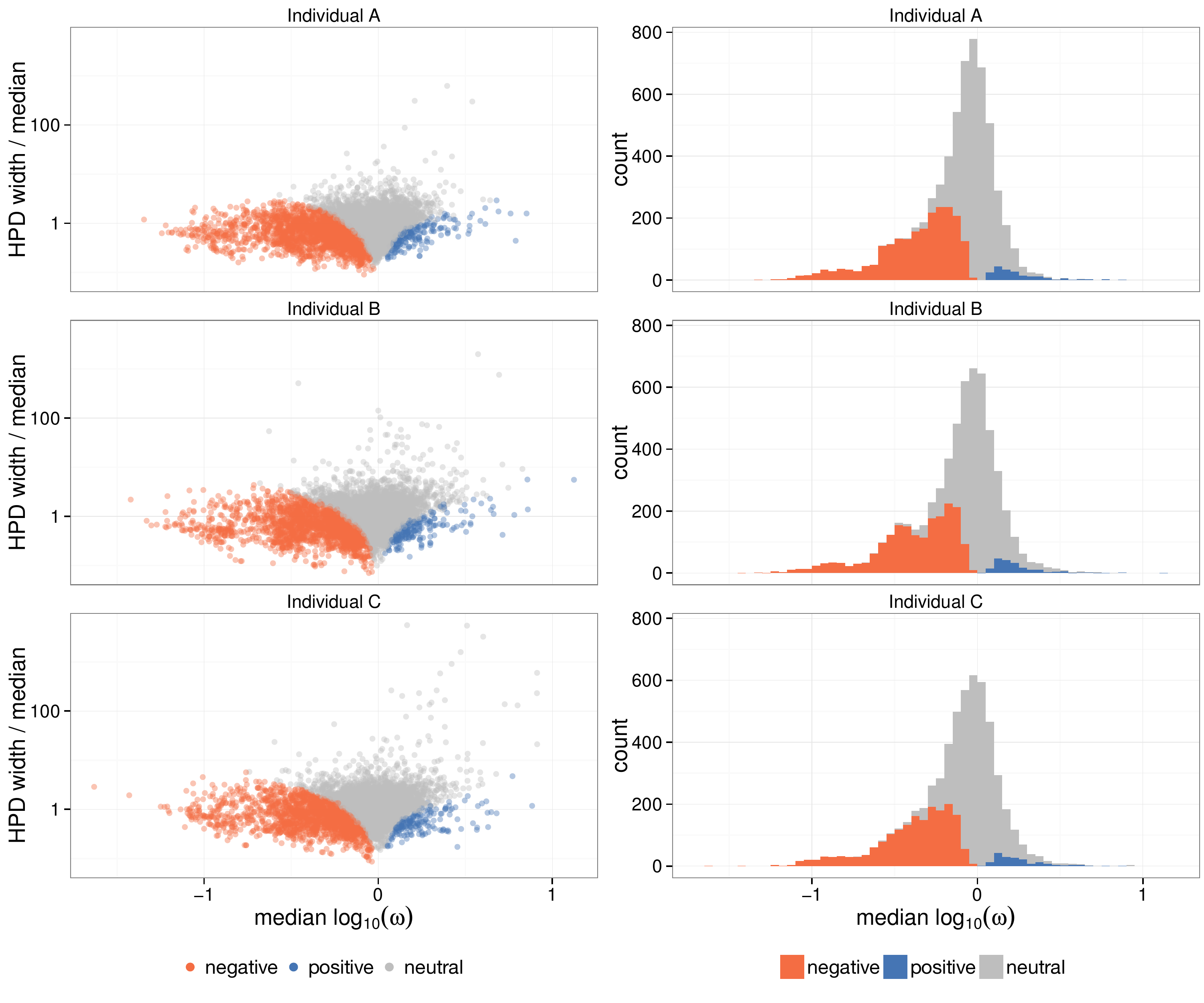}}
\end{center}
\caption{\
  Site-specific selection estimates partitioned by individual and gene.
  Sites classified as negatively selected or positively selected based on whether the 95\% Bayesian credible interval excludes 1 and in what direction.
  Left: comparison of $\omega$ estimate and relative width of BCI region.
  Right: distribution of site-specific selection estimates.
}
\label{FIGdndsDistribution}
\end{figure}
}

\newcommand{\FIGdndsBySite}{\
\begin{figure}
\begin{center}
  \arxiv{\includegraphics[width=0.8 \textwidth]{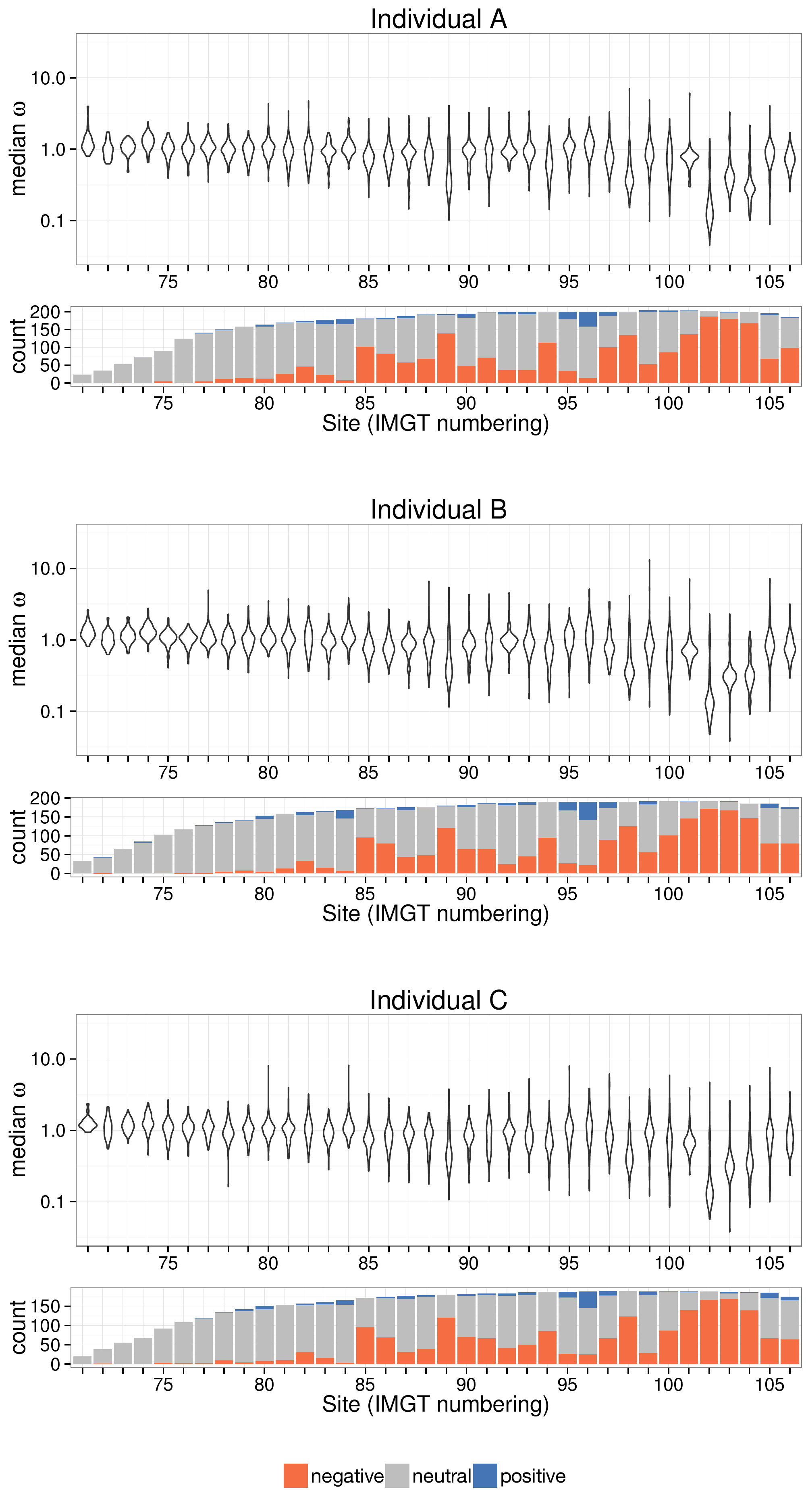}}
\end{center}
\caption{\
  Site-specific estimates of the selection coefficient $\omega$.
  Violin plots show distribution of median $\omega$ estimates across V genes at each site.
  Bar plots show count of V genes classified as undergoing negative, neutral, or positive selection.
  Only sites with at least 100 productive and out-of-frame observed sequences aligned were considered.
  The sites with IMGT numbers less than or equal to 104 are traditionally designated ``framework.''
}
\label{FIGdndsBySite}
\end{figure}
}

\newcommand{\FIGdndsConsistency}{\
\begin{figure}
\begin{center}
  \arxiv{\includegraphics[width=0.8 \textwidth]{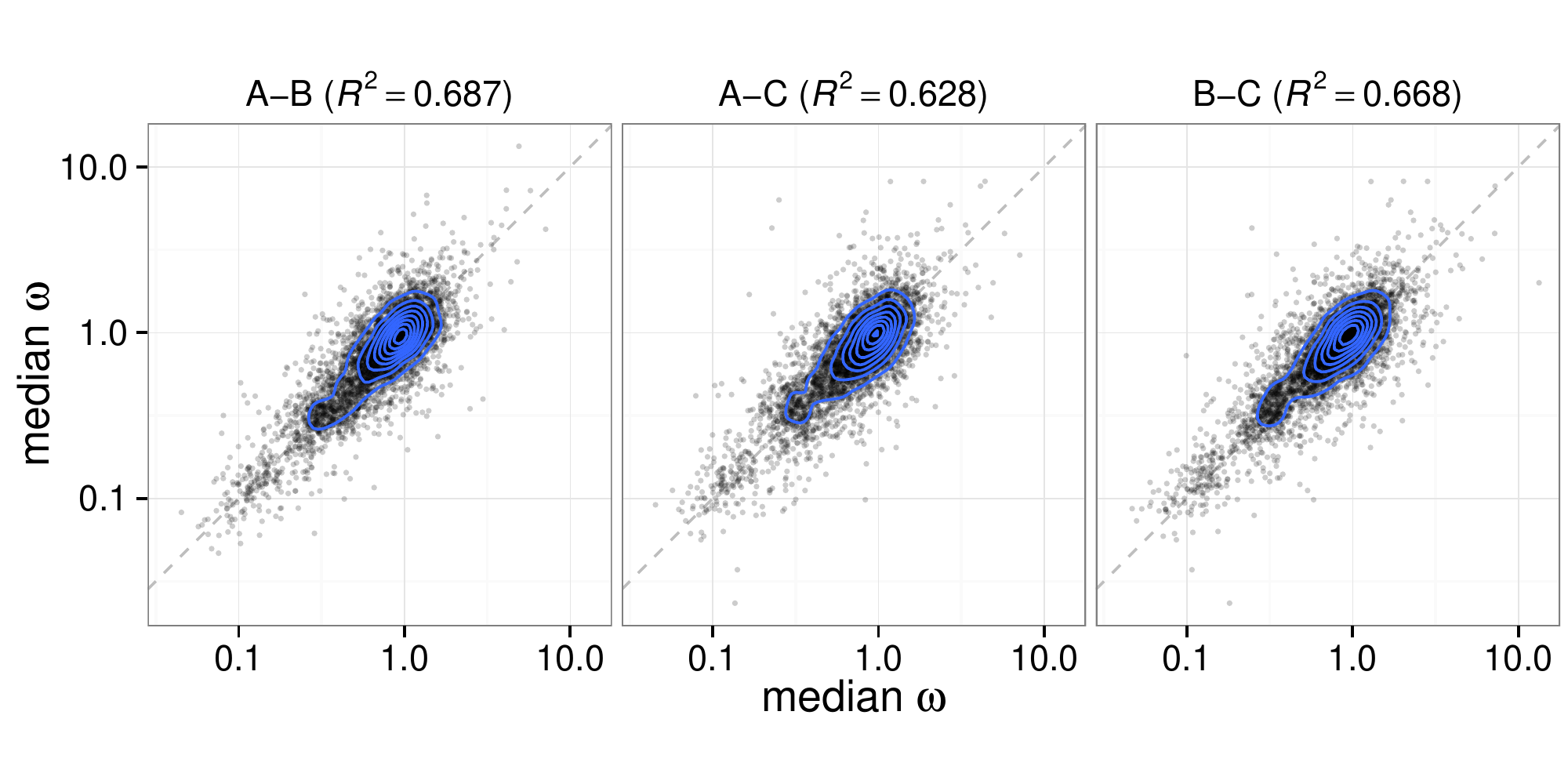}}
\end{center}
\caption{\
  Pairwise comparisons of site-specific $\omega$ estimates between the three individuals along with the $R^2$ value from a linear model fit using $\log_{10}(\omega)$ for both the predictor (x-axis) and response (y-axis).
}
\label{FIGdndsConsistency}
\end{figure}
}

\newcommand{\FIGgtrConsistency}{\
\begin{figure}
\begin{center}
  \arxiv{\includegraphics[width=\textwidth]{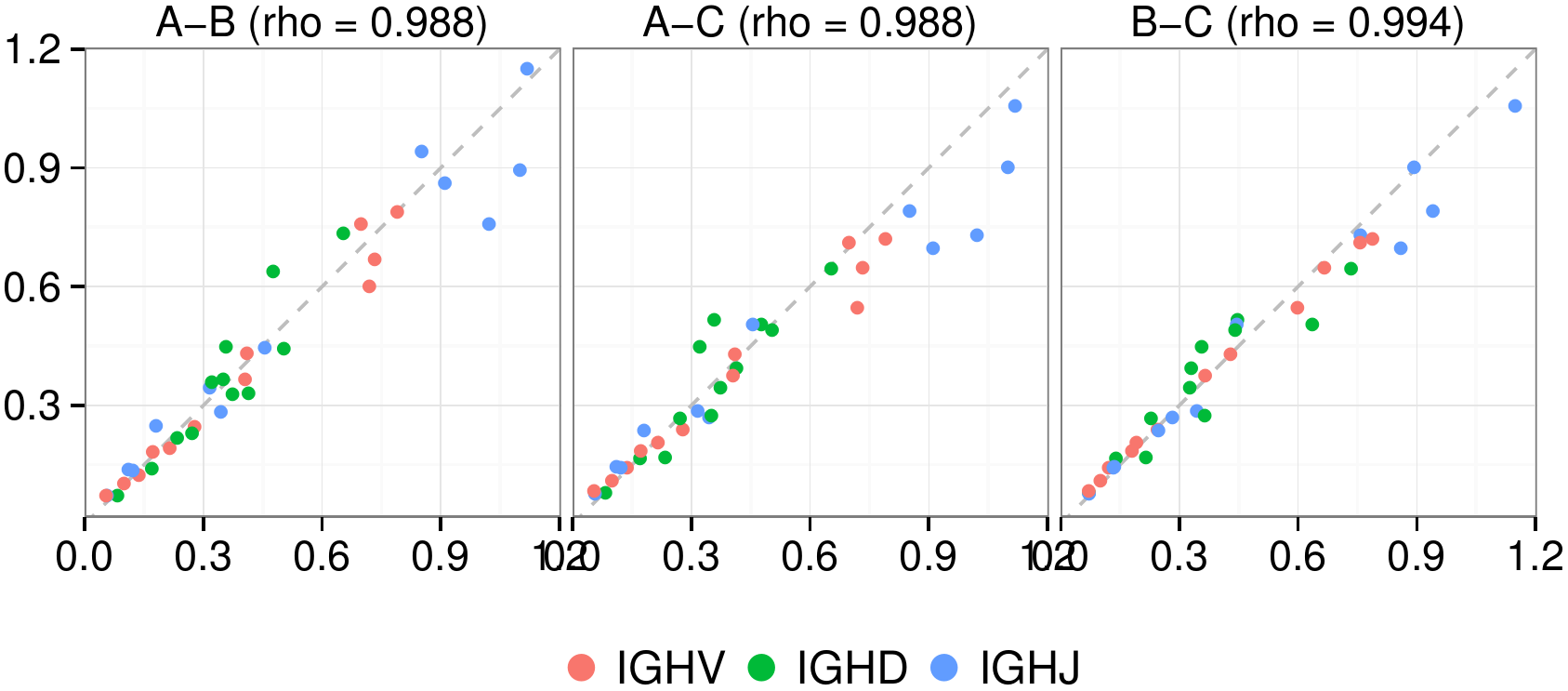}}
\end{center}
\caption{\
  Pairwise comparison of off-diagonal entries in maximum-likelihood $Q$ matrices under the \trqigi\ model between the three individuals.
  Coefficients are shown in Fig.~\ref{FIGgtr}.
}
\label{FIGgtrConsistency}
\end{figure}
}

\newcommand{\FIGsimulationResults}{\
\begin{figure}
\begin{center}
  \arxiv{\includegraphics[width=\textwidth]{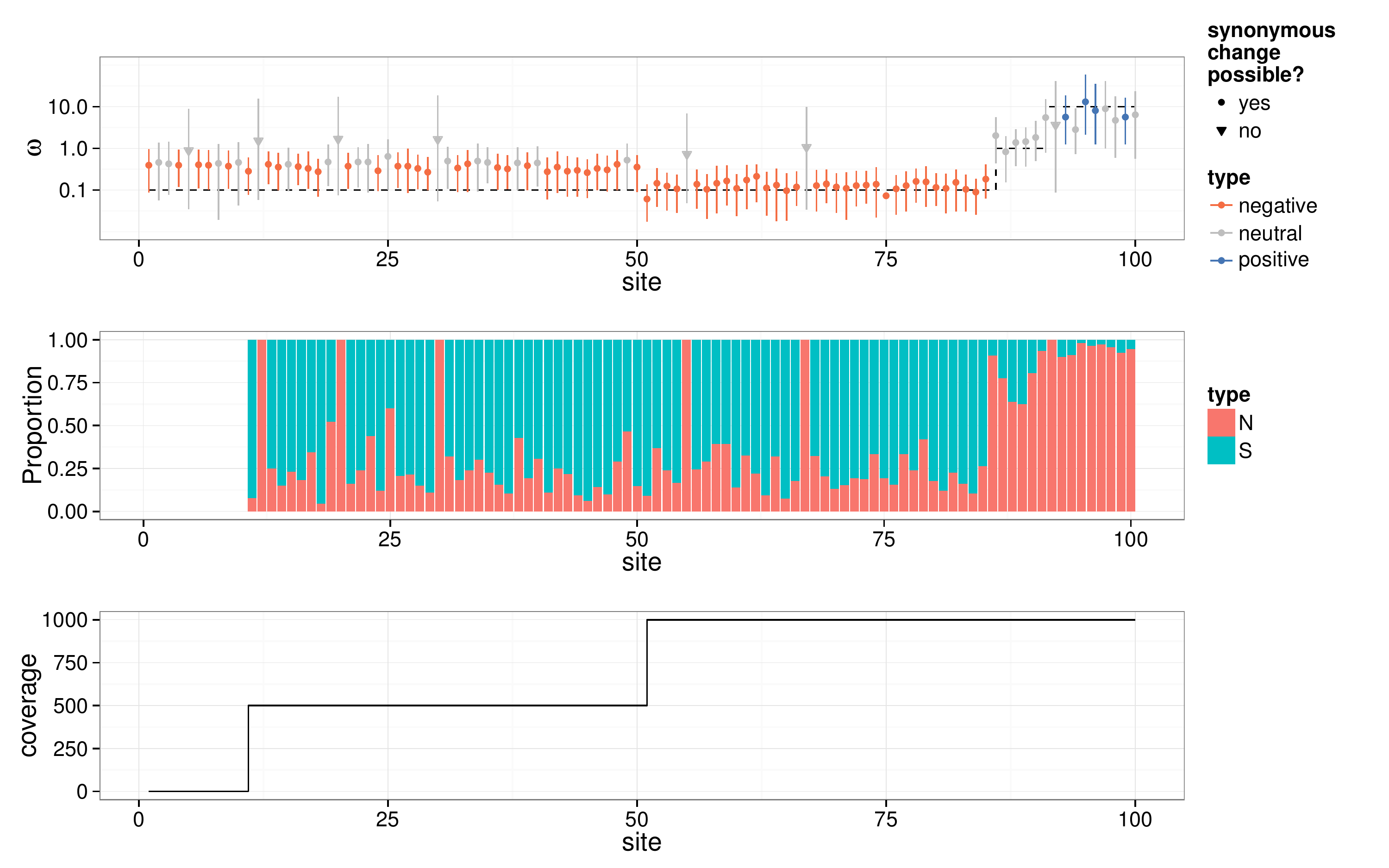}}
\end{center}
\caption{\
  Top panel: site-specific $\omega$ estimates under simulated data with varying coverage.
  Inverted trianges show sites where the germline state was Tryptophan or Methionine, from which no synonymous changes are possible.
  Dashed black line shows simulated $\omega$.
  Middle panel: proportion (second panel) of mutations at each position which were nonsynonymous (N) or synonymous (S).
  Bottom panel: sequence coverage by codon position.
}
\label{FIGsimulationResults}
\end{figure}
}

\newcommand{\FIGaa}{\
\begin{figure}
\begin{center}
  \arxiv{\includegraphics[width=\textwidth]{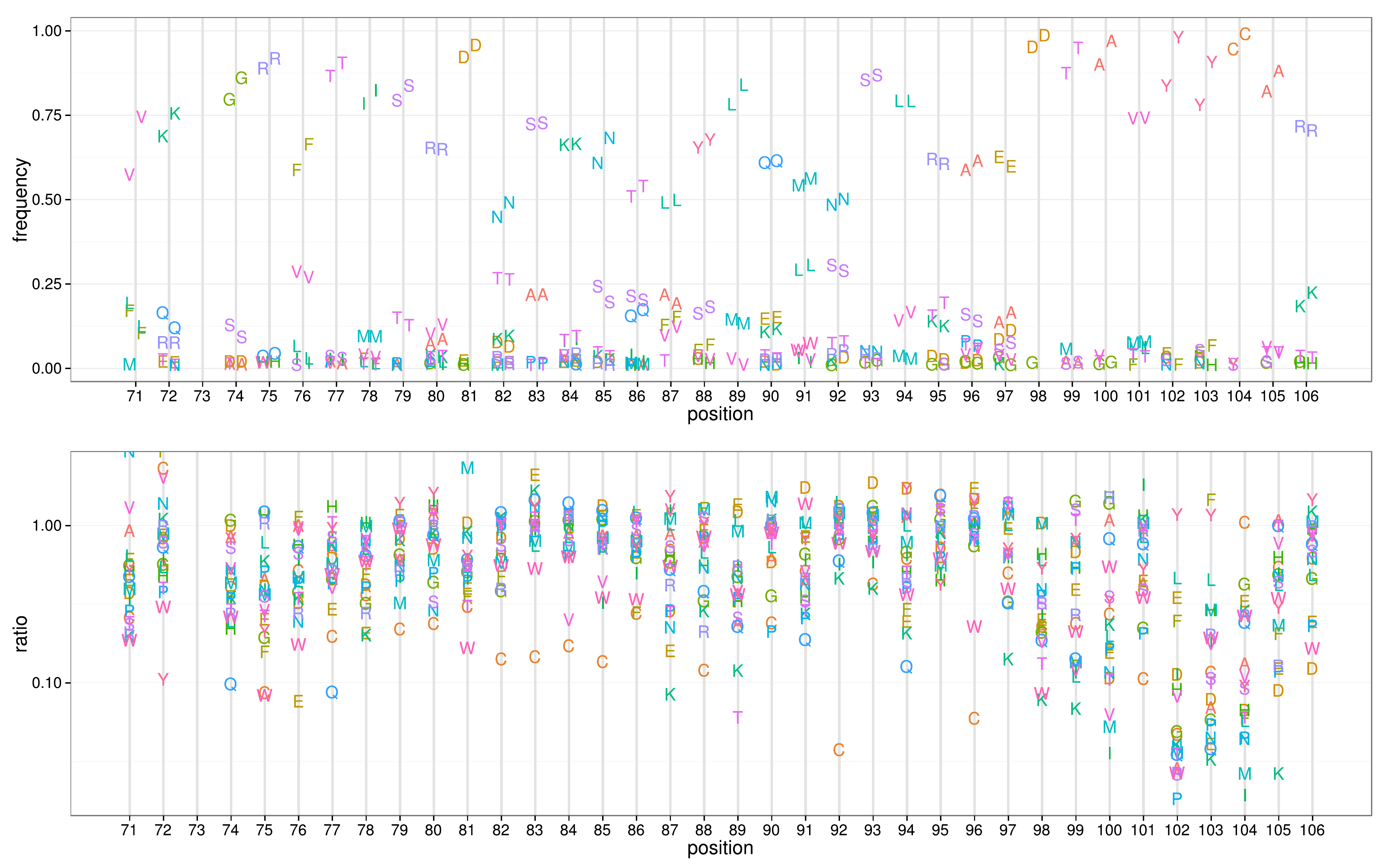}}
\end{center}
\caption{\
  Amino acid profiles of out-of-frame and functional B cell sequences as aligned by the IMGT alignment.
  Top panel: frequency of amino acids per site. Letters to the left of the line show the profile for out-of-frame sequences and those to the right of the line show the profile for functional sequences.
  Bottom panel: amino acid frequency in functional sequences divided by that in out-of-frame sequences.
}
\label{FIGaa}
\end{figure}
}

\newcommand{\FIGsasa}{\
\begin{figure}
\begin{center}
  \arxiv{\includegraphics[width=0.65 \textwidth]{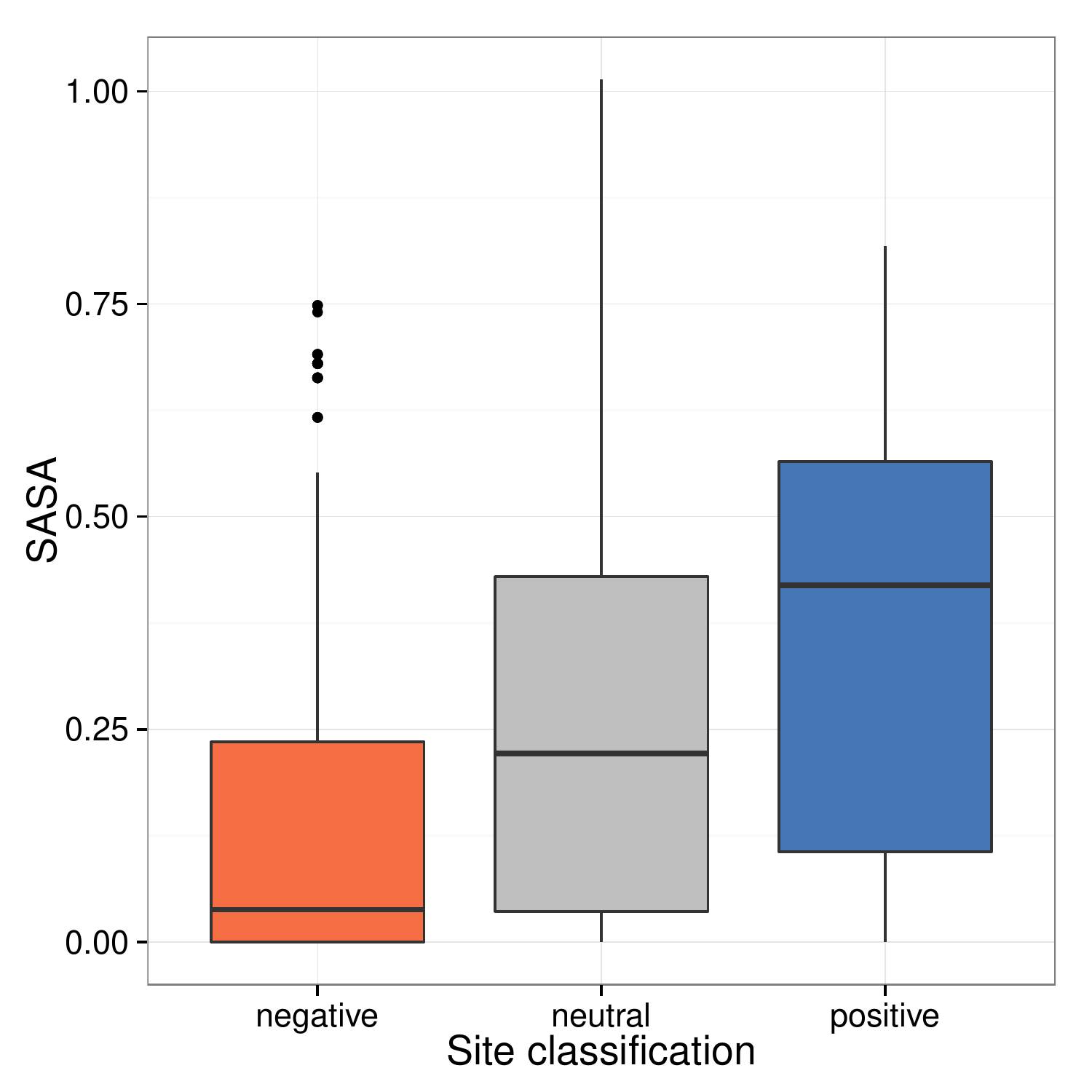}}
\end{center}
\caption{\
  Normalized solvent-accessible surface area (SASA) values by per-site $\omega$ classification.
  A SASA value of 1.0 indicates that the residue is fully exposed, while a value of 0.0 indicates that the residue is buried.
  Sites under negative selection are significantly less exposed than sites under positive selection ($p < 10^{-12}$) or neutral selection ($p < 10^{-15}$) by Bonferroni-corrected Wilcoxon rank-sum test.
  Neutral sites were less exposed than sites under positive selection ($p < 0.002$).
}
\label{FIGsasa}
\end{figure}
}

\newcommand{\FIGstructure}{\
\begin{figure}
\captionsetup[subfigure]{labelformat=empty}
\begin{center}
  \arxiv{\includegraphics[width=\textwidth]{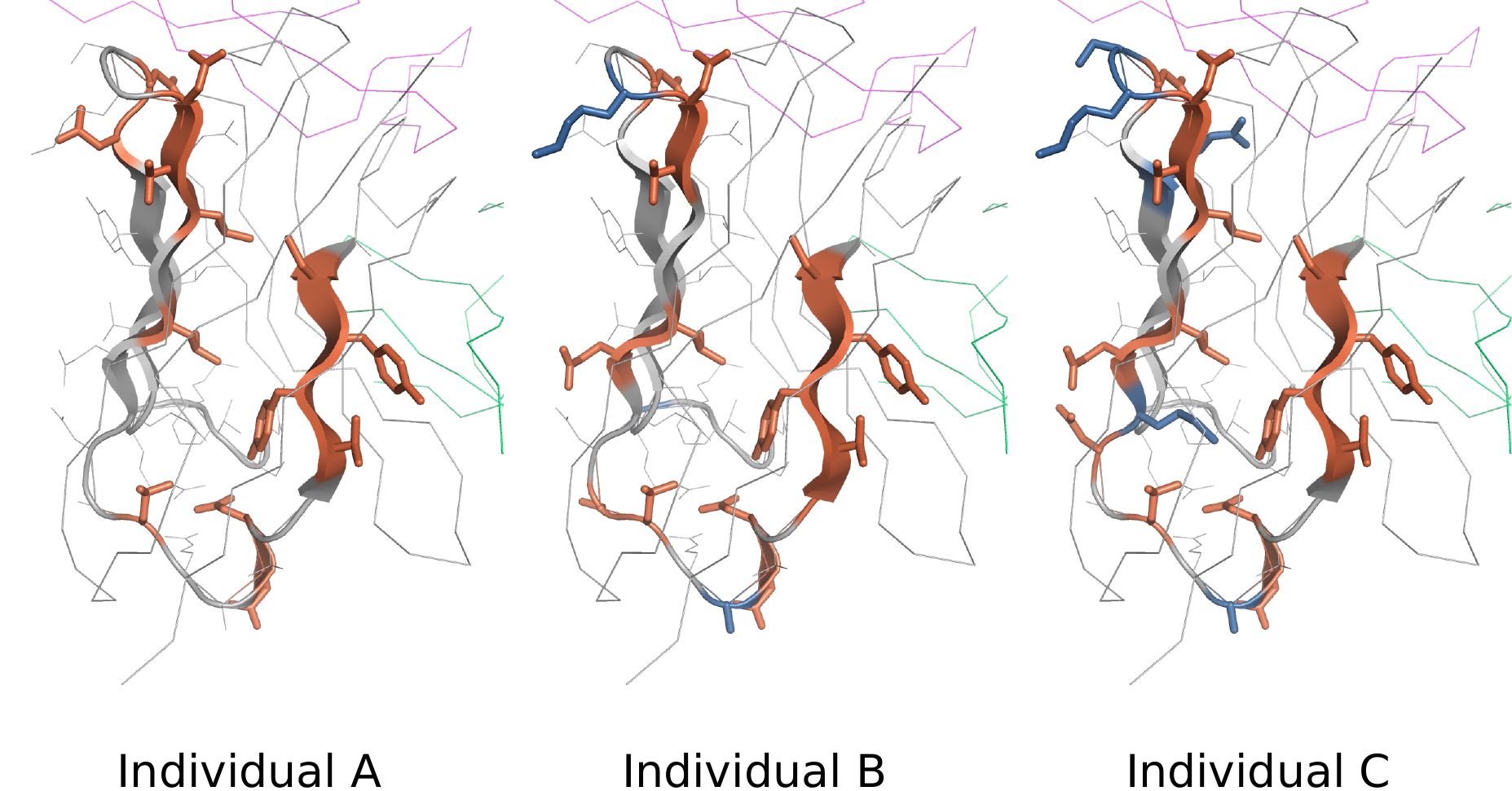}}
\end{center}
\caption{\
  An IGHV3-23*01 (the most frequent V gene/allele combination) heavy chain antibody in complex with IL-17A (PDB ID 2VXS;~\cite{Gerhardt2009-yo}), with sites colored by $\omega$ classification in each of the three individuals sampled.
  The bound antigen is shown in pink (top), and the light chain in green (right).
  The heavy chain structure is shown as a thin gray line at sites which could not be classified due to insufficient coverage.
  When there is sufficient coverage it is shown as a cartoon (thick lines or arrows representing beta sheets) which is colored gray at neutral sites, red at negatively selected sites, and blue at positively selected sites.
}
\label{FIGstructure}
\end{figure}
}

\newcommand{\tiqigi}{$t_i\, Q_i\, \Gamma_i$}
\newcommand{\trqigi}{$t_r\, Q_i\, \Gamma_i$}
\newcommand{\trqigs}{$t_r\, Q_i\, \Gamma_s$}
\newcommand{\trqsgs}{$t_r\, Q_s\, \Gamma_s$}

\newcommand{\TABLEmodel}{\
\begin{table}
\caption{Models and model testing results.}
\begin{subtable}{\textwidth}
\caption{The models of molecular evolution evaluated, including the number of free parameters introduced in parentheses.}\small
\label{TABLEmodelList}
\makebox[\textwidth][c]{\
  \begin{tabular}{l|p{1.5in}|p{1.5in}|p{1.5in}|p{0.7in}}
\hline
name						& branch length                                  								& GTR transition matrix						& across-site rate variation (discrete Gamma)									 & total parameters \\
\hline
\tiqigi		& One branch length per \emph{segment} per sequence ($n \times 3$)				& One matrix per segment ($8 \times 3$)		& One distribution per segment (3)								& $3n + 27$        \\
\hline
\trqigi		& One branch length per sequence ($n$) + relative rate between segments (2)		& One matrix per segment ($8 \times 3$)		& One distribution per segment (3)								& $n + 29$         \\
\hline
\trqigs		& One branch length per sequence ($n$) + relative rate between segments (2)		& One matrix per segment ($8 \times 3$)		& One shared distribution (1)									& $n + 27$         \\
\hline
\trqsgs		& One branch length per sequence ($n$) + relative rate between segments (2)		& One shared matrix (8)						& One shared distribution (1)									& $n + 11$ \\
\multicolumn{3}{c}{} 
\end{tabular}
}
\end{subtable}

\begin{subtable}{\textwidth}
\caption{Models show identical ranking across individuals. Columns include the log-likelihood (LogL), number of degrees of freedom (df), Akaike Information Criterion (AIC), and difference of AIC from the top model ($\Delta$AIC).}\small
\label{TABLEmodelTesting}
\centering
\begin{tabular}{llrrrr}
  \hline
    & model & LogL & df & AIC & $\Delta$AIC \\
  \hline
  A & \trqigi & -687,582 & 20,029 & 1,415,222 & 0 \\
    & \trqigs & -687,980 & 20,027 & 1,416,014 & 793 \\
    & \trqsgs & -700,818 & 20,009 & 1,441,654 & 26,433 \\
    & \tiqigi & -662,417 & 60,027 & 1,444,888 & 29,666 \\
   \hline
  B & \trqigi & -507,980 & 20,029 & 1,056,017 & 0 \\
    & \trqigs & -508,229 & 20,027 & 1,056,512 & 494 \\
    & \trqsgs & -517,320 & 20,009 & 1,074,658 & 18,641 \\
    & \tiqigi & -482,963 & 60,027 & 1,085,979 & 29,962 \\
   \hline
  C & \trqigi & -563,181 & 20,029 & 1,166,420 & 0 \\
    & \trqigs & -563,291 & 20,027 & 1,166,637 & 217 \\
    & \trqsgs & -572,530 & 20,009 & 1,185,078 & 18,659 \\
    & \tiqigi & -539,018 & 60,027 & 1,198,090 & 31,671 \\
   \hline
\end{tabular}
\end{subtable}

\label{TABLEmodel}
\end{table}\
}

\newcommand{\TABLEbaseFrequencies}{\
{\small
\begin{table}
\begin{tabular}{ll|llll|llll|llll}
          &          & \multicolumn{4}{|c|}{Individual A} & \multicolumn{4}{|c|}{Individual B} & \multicolumn{4}{|c}{Individual C} \\
          &          & A     &    G  &    C  &    T  &    A  &    G  &    C  &    T  &    A  &    G  &    C  &    T  \\
\hline
IGHV      & germline & 0.283 & 0.27  & 0.255 & 0.192 & 0.279 & 0.27  & 0.261 & 0.19  & 0.285 & 0.268 & 0.258 & 0.189 \\
          & sequence & 0.277 & 0.261 & 0.256 & 0.206 & 0.276 & 0.266 & 0.261 & 0.197 & 0.282 & 0.265 & 0.258 & 0.196 \\
\hline
IGHD      & germline & 0.199 & 0.328 & 0.141 & 0.332 & 0.196 & 0.323 & 0.157 & 0.324 & 0.197 & 0.326 & 0.153 & 0.324 \\
          & sequence & 0.197 & 0.315 & 0.168 & 0.321 & 0.198 & 0.309 & 0.176 & 0.317 & 0.197 & 0.314 & 0.172 & 0.317 \\
\hline
IGHJ      & germline & 0.197 & 0.428 & 0.22  & 0.154 & 0.2   & 0.424 & 0.223 & 0.154 & 0.186 & 0.438 & 0.225 & 0.151 \\
          & sequence & 0.186 & 0.433 & 0.222 & 0.159 & 0.193 & 0.427 & 0.224 & 0.156 & 0.18  & 0.44  & 0.227 & 0.153
\end{tabular}
\caption{Empirical stationary distribution for germline and observed sequences.}
\label{TABLEbaseFrequencies}
\end{table}}}

\newcommand{\TABLEreadCounts}{\
\begin{table}
\centering
\begin{tabular}{lrrr}
  \hline
  & \multicolumn{3}{c}{sequence count} \\
  individual & raw & unique by well & unique overall \\
  \hline
  A & 52,381,123   & 8,275,848          & 4,778,427 \\
  B & 59,241,547   & 9,820,657          & 5,826,068 \\
  C & 66,469,248   & 8,452,997          & 4,419,453 \\
  \hline
\end{tabular}
\caption{\
  Number of memory BCR sequences obtained by individual.
  ``raw'' refers to the number of reads obtained from sequencing, ``unique by well'' the number of unique sequences after performing clustering on reads for each barcoded PCR well, and ``unique overall'' the total number of unique sequences in the sample.
}
\label{TABLEreadCounts}
\end{table}}

\newcommand{\TABLEnumGermline}{\
\begin{table}
\begin{tabular}{llll}
individual         & cell type & in-frame & out-of-frame \\ \hline
\multirow{2}{*}{A} & na\"ive   & 0.93     & 0.92         \\
                   & memory    & 0.08     & 0.15         \\ \hline
\multirow{2}{*}{B} & na\"ive   & 0.91     & 0.89         \\
                   & memory    & 0.22     & 0.27         \\ \hline
\multirow{2}{*}{C} & na\"ive   & 0.92     & 0.90         \\
                   & memory    & 0.20     & 0.28         \\ \hline
\end{tabular}
\caption{\
  Fraction of BCR sequences that were identical to germline in the regions inferred to derive from germline.
  The fractions are stratified by individual, cell type, and frame status.
}
\label{TABLEnumGermline}
\end{table}}

\title{Quantifying evolutionary constraints on B cell affinity maturation}
\author[McCoy]{Connor O. McCoy}
\author[Bedford]{Trevor Bedford}
\author[Minin]{Vladimir N. Minin}
\author[Bradley]{Philip Bradley}
\author[Robins]{Harlan Robins}
\author[Matsen]{Frederick A. Matsen IV}
\date{\today}

\begin{document}

\begin{abstract}
The antibody repertoire of each individual is continuously updated by the evolutionary process of B cell receptor mutation and selection.
It has recently become possible to gain detailed information concerning this process through high-throughput sequencing.
Here, we develop modern statistical molecular evolution methods for the analysis of B cell sequence data, and then apply them to a very deep short-read data set of B cell receptors.
We find that the substitution process is conserved across individuals but varies significantly across gene segments.
We investigate selection on B cell receptors using a novel method that side-steps the difficulties encountered by previous work in differentiating between selection and motif-driven mutation;
this is done through stochastic mapping and empirical Bayes estimators that compare the evolution of in-frame and out-of-frame rearrangements.
We use this new method to derive a per-residue map of selection, which provides a more nuanced view of the constraints on framework and variable regions.
\end{abstract}

\maketitle


\section*{Keywords}
B cell, antibody, immunoglobulin, affinity maturation, molecular evolution, natural selection

\section*{Introduction}
Antibodies encoded by somatically modified human B cell receptor (BCR) genes bind a vast array of antigens, initiating an immune response or directly neutralizing their target.
This diversity is made possible by the processes of \textit{VDJ recombination}, in which random joining of V-, D-, and J-genes generates an initial combinatorial diversity of BCR sequences, and \textit{affinity maturation}, which further modifies these sequences.
The affinity maturation process, in which antibodies increase binding affinity for their cognate antigens, is essential to mounting a precise humoral immune response.
Affinity maturation proceeds via a nucleotide substitution process that combines Darwinian mutation and selection processes.
Mutational diversity is generated by \textit{somatic hypermutation} (SHM), in which a targeted molecular mechanism mutates the BCR sequence.
This diversity is then passed through a selective sieve in which B cells that bind well to antigen are stimulated to divide, while those that do not bind well or bind to self are marked for destruction.
The combination of VDJ recombination and affinity maturation enables B cells to respond to an almost limitless diversity of antigens.
Understanding the substitution process and selective forces shaping the diversity of the memory B cell repertoire thus has implications for disease prophylaxis and treatment.

It has recently become possible to gain detailed information about the B cell repertoire using high-throughput sequencing \citep{Boyd2009-ci,Wu2010-kx,Larimore2012-lo,DeKosky2013-rk,Robins2013-ww}.
Recent reviews have highlighted the need for new computational tools that make use of BCR sequence data to bring new insight, including the need for reproducible computational pipelines \citep{Mehr2012-se,Six2013-eo,Warren2013-fs,Georgiou2014-sh}.
Rigorous analysis of the B cell repertoire will require statistical analysis of how evolutionary processes define affinity maturation.
Statistical nucleotide molecular evolution models are often described in terms of three interrelated processes: mutation, the process generating diversity, selection, the process determining survival or loss of mutations, and substitution, the observed process of evolution that follows from the first two processes.
One major vein of research has focused on how nucleotide mutation rates depend on the identity of surrounding nucleotides (reviewed in \citet{Delker2009-xb}; see also \citet{Yaari2013-dg,Kepler2014-jy}), but little has been done concerning other aspects of the process, such as how the substitution process differs between gene segments.

Along with mutation, selection due to competition for antigen binding forms the other key part of the affinity maturation process.
Inference of selective pressures in this context is complicated by nucleotide context-dependent mutation, leading some authors to proclaim that such selection inference is not possible \citep{Dunn-Walters1998-ra}.
Indeed, if one does not correct for context-dependent mutation bias, interactions between those motifs and the genetic code can lead to false positive identification of selective pressure.
Previous work has developed methodology to analyze selection on sequence tracts in this context (reviewed in Discussion), but no methods have yet achieved the goal of statistical per-residue selection estimates.
This has, however, been recently identified as an important goal \citep{Yaari2013-dg}.
Such selection estimates could be used to better direct generation of synthetic libraries of antibodies for high throughput screening.
Another application would be to the engineering of antibody Fc regions with specific properties, such as for bispecific monoclonal antibodies or antibody-derived fragments, while preserving overall stability.

The ensemble of germline V, D, and J genes that rearrange to encode antibodies (equivalently: immunoglobulins) are divided into nested sets.
They can first be identified by their \textit{locus}: IGH, denoting the heavy chain, IGK, denoting the kappa light chain, or IGL, denoting the lambda light chain.
Our dataset contains solely the IGH locus, so we will frequently omit the locus prefix for simplicity.
Genes within a locus can be first subdivided by their \textit{segment}, which is whether they are a V, D, or J gene.
IGHV genes are further divided into \textit{subgroups} which share at least 75\% nucleotide identity.
Genes also have polymorphisms that are grouped into \textit{alleles}, which represent polymorphisms of the gene between individuals \citep{Lefranc2008-kg}.

VDJ recombination does not always produce a functional antibody, such as when the V and J segments are not in the same reading frame after recombination (an \textit{out-of-frame} rearrangement) or when the receptor sequence contains a premature stop codon.
However, each B cell carries two copies of the IGH locus, with one on each chromosome.
If the rearrangement on the first locus fails to produce a viable antibody, the~second locus will rearrange; if this second rearrangement is successful, the antibody encoded by the second rearrangement will be produced by the cell \citep{Corcoran2005-gp}.
If this second rearrangement does not produce a viable antibody the cell dies.

When assaying the BCR repertoire through sequencing, some of the sequences will be from cells for which the first rearrangement was successful, while others will be from cells with one productive and one out-of-frame rearrangement.
Although the out-of-frame rearrangements from the second type of cell do not produce viable antibodies, their DNA gets sequenced along with the productive rearrangements.
Since SHM rarely introduces insertions or deletions (we observe whole codon insertion deletion events in between 0.013\% to 0.014\% of memory sequences within templated segments), it is appropriate to assume that observed frame shifts in sequences are dominated by out-of-frame rearrangement events.
However, because they are not expressed, but rather are carried along in cells with a separate functional rearrangement, they have no selective constraints.
For this reason, we use sequences from out-of-frame rearrangements as a proxy for the neutral mutation process in affinity maturation.

In this paper, we develop modern statistical molecular evolution methods for the analysis of high-throughput B cell sequence data, and then apply them to a very deep short-read data set of B cell receptors.
Specifically, we first apply model selection criteria to identify patterns in the single-nucleotide substitution process that occurs during affinity maturation and find that they are similar across individuals but vary significantly across gene segments.
Next, we investigate how substitution processes vary between V genes and find that the primary source of variation is whether or not a sequence produces a functional receptor.
Finally, we develop the first statistical methodology and corresponding software for comprehensive per-residue selection estimates for B cell receptors.
We leverage out-of-frame rearrangements carried along in B cells with a productively rearranged receptor on the second chromosome to estimate evolutionary rates under neutrality, thus avoiding difficulties encountered by previous work in differentiating between selection and motif-driven mutation.
A key part of our method is our extension of the ``counting renaissance'' method for selection inference \citep{Lemey2012-ja} for non-constant sequencing coverage and a star-tree phylogeny.
Using this modified method, we are able to efficiently derive a per-residue map of selection on more than 15 million B cell receptor sequences; we find that selection is dominated by negative selection with patterns that are consistent among individuals in our study.

\section*{Results}

\subsection*{Substitution model inference and testing}
We evaluated the fit of nested models with varying complexity, ranging from a simple model with shared branch lengths and substitution processes for the three independent segments of the BCR, to a complex model with completely separate substitution processes and branch lengths for each segment (Tab.~\ref{TABLEmodelList}).
For the underlying nucleotide substitution model, we fit a general time-reversible (GTR) nucleotide model \citep{Lanave1984-mf} with instantaneous rate matrix $Q$ to subsets of the data, using 20,000 unique sequences from each individual.
The choice of a stationary and reversible model, rather than a more general model, was based on the similarity of base frequencies between the germline and observed sequences (Tab.~\ref{TABLEbaseFrequencies}).
We modeled substitution rate heterogeneity across sites using a four-category discretized Gamma distribution \citep{Yang1994-ik} with fixed mean 1.0.

\arxiv{\TABLEmodel}
We find that the best performing model (denoted \trqigi, Tab.~\ref{TABLEmodelTesting}) is one in which the branch length separating a sequenced BCR from its germline counterpart is estimated independently for each observed sequence, but that V, D and J regions differ systematically in their relative amounts of sequence change (denoted $t_r$).
Additionally, this model uses separate GTR transition matrices for V, D and J regions (denoted $Q_i$) and uses separate distributions for across-site rate variation for V, D and J regions (denoted $\Gamma_i$).
Looking across models, both the Akaike Information Criterion (AIC)\citep{Akaike1974-pz} (Tab.~\ref{TABLEmodelTesting}) and the Bayesian Information Criterion \citep{Schwarz1978-he} (data not shown) identified the same rank order of support; this ordering was also identical for each of the three individuals.
Other than the \tiqigi\ model, in which branch length is estimated independently across gene segments, models are ranked in terms of decreasing complexity.
The finding that a complex model fits better than simpler models is likely aided by the large volume of sequence data available.

Next, we fit the best-scoring model (\trqigi) to the full data set for each individual.
The median distance to germline was 0.063, 0.030, and 0.039 substitutions per site for individuals A, B, and C, respectively.
The distribution of branch lengths appears nearly exponential for individuals B and C, with many sequences close to germline and few distant from germline sequences (Fig.~\ref{FIGgtrBl}).
Individual A displayed a higher substitution load and a non-zero mode.
Despite these differences in evolutionary distance, the relative rate of substitution between the V, D, and J segments for each individual was very similar.
We note that the sorting procedure used to separate memory from na\"{i}ve B cells provided memory cells at approximately 97\% purity, so these divergence estimates may be conservative due to low levels of contamination from the na\"{i}ve repertoire.

\arxiv{\FIGgtrBl}
Coefficients from the GTR models for the same gene segment across individuals were quite similar to one another, while models for different gene segments within an individual showed striking differences (Fig.~\ref{FIGgtr},~\ref{FIGgtrConsistency}).
However, overall correlations of GTR parameters between individuals were very high, yielding correlation coefficients between $\rho = 0.988$ and $\rho = 0.994$.
We observe an enrichment of transitions relative to transversions in all segments, as previously described \citep{Teng2007-vx}.

Next we compared the evolutionary process between various groupings of sequences to learn what determines the characteristics of this evolutionary process.
We focused on the V gene segment, as it had the most coverage in our dataset, and partitioned the sequences by whether they were in-frame, then by individual, and then by gene subgroup.
We fit the \trqigi\ model to 1000 sequences from each set of the partition and calculated the transition probability matrix ($P$) associated with the median branch length across all sequences given an equiprobable starting state.
These matrices were then analyzed with a variant of compositional principal components analysis \citep{Aitchison1983-if} (see Materials and Methods).
We find that substitution models are influenced by in- versus out-of-frame sequence status, find no evidence for models clustering by individual, and see some limited evidence for clustering by gene subgroup (Fig.~\ref{FIGcompPca}).
The Euclidean distance between these transformed discrete probability distributions and the Hamming distance between germline V genes showed significant, but moderate, correlation (Spearman's $\rho = 0.20$, $p < 10^{-15}$; Fig.~\ref{FIGmodelVsHammingDist}).

\arxiv{\FIGcompPca}

\subsection*{Natural selection}

The primary challenge for B cell receptor selection inference is that nucleotide context is known to have a very strong impact on mutation rates (reviewed in \citep{Teng2007-vx}).
These context-specific mutations combined with the structure of the genetic code can result in extreme $dN/dS$ ratios using the classical definition that are not attributable to selection.
To address this problem, we infer the selection coefficient $\omega$ using a nonsynonymous-synonymous ratio which controls for background mutation rate via out-of-frame sequences \eqref{eq:omega}.
We continue the tradition of calling the selection coefficient $\omega$ in this context, even though it is a slightly different definition than previously used.

Applying this method to our data set results in the first per-site and per-gene maps quantifying selection in the B cell repertoire \cite{figshare-plots,figshare-csv}.
Sites were classified as negatively or positively selected based on whether the 95\% Bayesian credible interval (BCI) excludes 1.0: sites for which the lower endpoint of the $\omega$ BCI is greater than 1.0 are classified as being under positive selection, while sites for which the upper endpoint of their $\omega$ BCI is less than 1.0 are classified as being under negative selection.
We employ site numbering according to the IMGT unique numbering for the V domain \citep{Lefranc2003-xk}.

\arxiv{\FIGdndsNsExample}
IGHV3-23*01 is the most frequent V gene/allele combination in our dataset, and it displays patterns that are consistent with the other genes.
Specifically, we see significant variation in the synonymous substitution rate (right panels, Fig.~\ref{FIGdndsNsExample}a) even in out-of-frame sequences, which is presumably due to motif-driven mutation.
Thus, if we had directly applied traditional means of estimating selection by comparing the rate of nonsynonymous and synonymous substitutions, we would have falsely identified sites as being under strong selection.
In contrast the selection inferences made using out-of-frame sequences stay much closer to neutral (Fig.~\ref{FIGdndsNsExample}b).

\arxiv{\FIGdndsBySite}
We note extensive negative selection in the residues immediately preceding the CDR3 (Fig.~\ref{FIGdndsBySite}).
The amino acid profile for these sites shows a distinct preference for a tyrosine or rarely a phenylalanine two residues before the start of the CDR3 at site 102 (Fig.~\ref{FIGaa}).
It shows a preference for a tyrosine or more rarely a phenylalanine or a histidine in the residue just before the start of the CDR3 at site 103.
These aromatic positions likely play important structural roles in the antibody complex: site 102 is buried in the core of the heavy chain and makes extensive van der Waals interactions as well as a sidechain-backbone hydrogen bond, while site 103 forms part of the interface between the heavy and light chains (see further description of structural results below).

\arxiv{\FIGdndsDistribution}
Overall we see extensive selection in our sequenced region (Fig.~\ref{FIGdndsDistribution}).
The mean $\omega$ estimate across sites with at least 100 productive and out-of-frame sequences aligned was 0.907.
65.6\% of sites had a median $\omega < 1$ with a wide distribution of median $\omega$ values and confidence interval widths.
However, many of them were observed to be positively, negatively, and neutrally evolving with narrow confidence intervals (Fig.~\ref{FIGdndsDistribution}, left column).
30.6\% of sites were confidently classified as being under negative selection (Fig.~\ref{FIGdndsDistribution}, right column).

Because amino acids interior to the protein could be important for protein stability compared to exposed ones, we hypothesized that residues under negative selection would be more internal to the antibody protein than those under neutral or positive selection, and that the inverse would be true for residues under positive selection.
To test this, we mapped our $\omega$ estimates onto antibody structures (Fig.~\ref{FIGstructure}) and calculated the exposure of each amino acid position in the structure using the solvent-accessible surface area (SASA) using ROSETTA3~\citep{Leaver-Fay2011-my}.
The normalized SASA was well correlated with the classification of each site: sites classified as being under positive selection were most exposed in the protein structure, followed by neutral sites, then negatively selected sites (Fig.~\ref{FIGsasa}).
Differences in surface accessibility were significant between the three groups, with p-values ranging from $<0.002$ for the comparison of positive vs.\ neutral sites to $<10^{-15}$ for the comparison of negative vs.\ neutral sites (Wilcoxon rank-sum test~\citep{hollander1973nonparametric}).

\arxiv{\FIGstructure}
Despite the three individuals surveyed here presumably having quite different immune histories, we observe remarkable consistency in substitution and selection within the memory B cell repertoire.
Indeed, we see a very strong correlation of median selection estimates between individuals (Fig.~\ref{FIGdndsConsistency}), with between-individual coefficients of determination $R^2$ of between 0.628 and 0.687 for site-specific $\omega$ values.

\section*{Discussion}
We find different patterns of substitution across the V, D and J regions which is consistent among individuals (Fig.~\ref{FIGgtr}) even though those individuals have differing levels of substitution (Fig.~\ref{FIGgtrBl}).
We find that the dominant factor determining the V segment substitution process is whether it is out-of-frame or productive, with the gene identity being a contributing factor.
The pattern of selective pressure is consistent across individuals, and shows especially strong pressure near the boundary between the V gene and the CDR3.
Selection estimates for BCRs are still high, with average $\omega$ of $\approx 0.9$, compared to common examples of Darwinian evolution, such as seen in \textit{Drosophila} \citep{Clark2007} and mammals \citep{Lindblad2011}, where most genes show $\omega$ less than 0.1.
However, we note that although our estimates of $\omega$ are comparable to more traditional estimates, we calculate $\omega$ slightly differently, using out-of-frame sequences as a control for motif-driven evolution.
Finally, the patterns of selective pressure we observed correlated with levels of surface exposure in published antibody structures: highly conserved sites were more frequently found internally, while residues we classified as positively selected were more exposed.

We note that our analyses are based on data from only three individuals.
It is possible that including more individuals would reveal variation in the mutation process.
However, we note that these three unrelated individuals had an extraordinary level of agreement, which cannot be explained by sequencing error.

\subsection*{Substitution process}
We closely examine the substitution and selection processes in a context-independent manner, not to make a full description of this clearly context-dependent process, but rather to provide a solid framework for future study and to enable downstream comparative analyses (Figure~\ref{FIGcompPca}).
Our model selection shows that the best-fitting model allows for a single branch length per sequence, a global multiplier for per-segment differences, a per-segment substitution model, and a per-segment rate variation model across sites (Tab.~\ref{TABLEmodelTesting}).
These between-segment differences are certainly due in part to base composition, which also differs significantly between segments and is similar between individuals (Table~\ref{TABLEbaseFrequencies}).
Another contributing factor is probably similarity of local nucleotide context between the genes of a given segment compared to between segments; these nucleotide contexts are known to impact AID-induced somatic hypermutation (reviewed in \citet{Teng2007-vx}).
We also note that the entirety of the D segment lies within the CDR3 region, and is thus more likely to directly contact an epitope; not surprisingly, we observe higher substitution rates within that segment.
By analyzing distances between GTR substitution rate matrices, we find that the most important difference between them is determined by whether they are productive or non-productive (Fig.~\ref{FIGcompPca}), which is presumably due to the impact of natural selection.
We also find a significant correlation between sequence identity and substitution matrix (compare \cite{Kosakovsky_Pond2010-ta}).
In a related though distinct vein, \citet{Mirsky2014-wu} develop an amino acid substitution model for BCR sequences, which analogously aggregates information across positions.

\subsection*{Selection process}
The role of selection in B cell receptor development has stimulated continuous interest since the pioneering 1985 paper of Clarke and colleagues \citep{Clarke1985-qx}, however methods for the analysis of antigen selection have developed in parallel to related work in the population genetics and molecular evolution community.
Work on the selection process for BCRs has focused on aggregate statistics to infer selection for entire sequences or sequence tracts, and there has been a lively debate about the relative merits of these tests
\citep{Chang1994-aj,Lossos2000-mq,Bose2005-gs,Hershberg2008-rp,Yaari2012-kk}.
Recent work has offered methods that evaluate selection on a per-sequence basis \citep{Yaari2012-kk}. 
There have also been efforts to infer selection based on lineage shape \citep{Steiman-Shimony2006-fm,Abraham2006-fd,Barak2008-fw,Shahaf2008-cc,Liberman2013-ur,Uduman2014-pb}, which has been a common approach in macroevolutionary studies (reviewed in \citet{Mooers1997-jl}) and more recently in population genetics \citep{Drummond2008-bl,Li2013-lg,Luksza2014-xf,Neher2014-vl}.

In this work, we develop the first means of inferring per-residue selection using high-throughput sequence data with non-uniform coverage.
Our method side-steps the difficulties encountered by previous work in differentiating between selection and motif-driven mutation in B cell receptors
\citep{Chang1994-aj,Dunn-Walters1998-ra,Lossos2000-mq,Bose2005-gs,Hershberg2008-rp,Yaari2012-kk,Yaari2013-dg}
by developing statistical means to compare in-frame and out-of-frame rearrangements.
An alternate means of estimating selection was recently developed by \cite{Kepler2014-jy} in which a regression model incorporating a detailed model of motif preferences was used to infer selection coefficients for the framework and CDR regions in aggregate.
In contrast to this previous work on B cell selection, our methods provide a \emph{per-residue} selection map for a contiguous stretch of BCR sequence.

We use out-of-frame rearrangements as our selection-free control population.
These sequences do not create functional IGH proteins, but may be carried in heterozygous B cells which do have a productively rearranged IGH allele.
Thus they undergo SHM, but any selection occurs on the level of the productively rearranged allele, not on the residues in the unproductive allele.
We observe a similarly high proportion of germline-identical sequences for in-frame and out-of-frame subsets in na\"ive cells (Table~\ref{TABLEnumGermline}); differences from germline derive in part from sequencing and other platform errors that do not depend on frame.
For memory cells, we see extensive action of somatic hypermutation, but with a higher proportion of germline-identical out-of-frame sequences than in-frame (Table~\ref{TABLEnumGermline}).
We interpret these additional mutations for in-frame memory sequences as following from the process of affinity maturation for a specific antigen.
We acknowledge that some out-of-frame sequences could still feel the impact of selection, which would occur if the sequences accrue frameshift mutations in the process of affinity maturation.
However, it is thought that SHM is primarily a process of point mutation \citep{Teng2007-vx}, and indeed, we observe whole codon indels in only 0.013\%--0.014\% of memory sequences within templated segments.
Still, if a weaker version of selection was occurring on the out-of-frame sequences compared to the productive ones then this would simply make our estimates of selection conservative, pulling estimates of $\omega$ closer to 1, and yet our selection estimates are confidently classified as non-neutral for a substantial fraction of sites (Fig.~\ref{FIGdndsDistribution}).

In applying our methodology to IGHV sequences, we gain a high resolution per-gene map of selective forces on B cell receptors for part of the V gene.
This part is primarily in the framework region, which is thought to be under substantial evolutionary constraint to preserve structure.
Indeed, we see an pattern of quite strong negative selection in the region around the beginning of the CDR3 (Figure~\ref{FIGdndsBySite}), agreeing with recent work that found strong negative selection in one site near the beginning of the CDR3 \citep{Yaari2013-dg}.
However, other sites in this section of framework have substantially relaxed selection (Figure~\ref{FIGdndsBySite}).
These results thus provide a more nuanced view into the constraints on B cell receptor sequences rather than the traditional framework/variable designations, as also noted by \citep{Yaari2013-dg}.

This work points the way towards future directions.
In this work we assumed that the size of individual lineages is small compared to the size of the overall repertoire, and thus that lineage structure could be ignored for the purpose of evolutionary model analysis.
Ideally we would reconstruct lineages and then do evolutionary analysis on the tree corresponding to each lineage.
However, reconstructing groups of sequences forming a lineage is a challenging prospect on its own, to say nothing of doing phylogenetics on sequences in the presence of strong context-specific mutation-selection patterns, and have left out incorporating those aspects until we have first developed the necessary methods.
We have recently developed an HMM framework to analyze VDJ rearrangements \cite{Ralph2015-kr}, and are currently developing and validating ways to use this framework for likelihood-based (as opposed to procedure-based \cite{Liao2013-cr,Bashford-Rogers2013-uy}) lineage group inference.

\section*{Materials and Methods}

\arxiv{\FIGoverview}

\subsection*{Data set}
The complete description of the experiment will be published separately (manuscript in preparation).
However, here we give a brief overview of the data in order to facilitate understanding of our analysis and to emphasize that the experimental design has a number of features that greatly reduce errors in sequencing and quantification.
400cc of blood was drawn from three healthy volunteers under IRB protocol at the Fred Hutchinson Cancer Research Center.
CD19$^+$ cells were obtained by bead purification then flow sorted to isolate over 10 million na\"ive (CD19$^+$CD27$^-$IgD$^+$IgM$^+$) and over 10 million memory (CD19$^+$CD27$^+$) B cells, with greater than 97\% purity.
Genomic DNA was extracted and the ImmunoSeq assay described in \citep{Larimore2012-lo} was performed on the six samples at Adaptive Biotechnologies in Seattle, WA\@.

The experiments and preprocessing were carefully designed to give an accurate quantification of error-corrected observed sequences.
To mitigate preferential amplification of some V/J pairs through primer bias, the PCR amplification was performed using primers optimized via a large collection of synthetic templates~\citep{Carlson2013-kx}.
To reduce sequencing errors and provide accurate quantification, each sample was divided amongst the wells on two 96 well plates and bar-coded individually.
These templates were then amplified and ``over-sequenced'' (Table~\ref{TABLEreadCounts}), so that an average of almost 6 reads were present for each template.
Following \citet{Robins2009-pe}, reads matching the same template were collapsed into a consensus sequence with reduced sequencing error.
Here, we grouped reads from within a well into consensus sequences by joining reads with Hamming distance less than or equal to two, and inferred the consensus sequence in each group using parsimony.
Groups with only one member were discarded.
This procedure protects against collapsing distinct sequences, as the probability that nearly identical distinct sequences co-occur exclusively in the same wells is small.
We acknowledge this procedure may eliminate low frequency variants, but we prioritized accuracy over sensitivity towards these variants; note that despite this conservative analysis pipeline we observed substantial signal in the data.

Deep sequencing these B cell populations resulted in 15,023,951 (Tab.~\ref{TABLEreadCounts}) unique 130bp observed sequences after pre-processing that spanned the third heavy chain complementarity determining region (CDR3) region (Fig.~\ref{FIGoverview}).
The full data set will be made public upon publication of the manuscript describing the experiment.

\subsection*{Alignment and germline assignment}
Each sequence was first aligned to each V gene using the Smith-Waterman algorithm with an affine gap penalty \citep{Gotoh1982-ff}.
The 3' portion of the sequence not included in the best V gene alignment was next aligned to all D and J genes available from the IMGT database \citep{Lefranc2008-kg}.
The best scoring V, D, and J alignment for each sequence was taken to be the germline alignment, and the corresponding germline sequence was taken to be the ancestral sequence for that observed sequence; in the case of ties, one germline sequence was chosen randomly among those alleles present at abundance $\ge 10\%$.
Sequences were classified as productive or out-of-frame based on whether the V and J segments were in the same frame; all sequences with stop codons were removed, as these sequences could result from either an unproductive rearrangement event or inactivation due to a lethal mutation.
The 18 V gene polymorphisms present at the highest frequency in the na\"{i}ve populations of the individuals surveyed which were not represented in the IMGT database were added to the list of candidates for alignment.
In contrast to na\"ive sequences which have no mutations across almost all sites, the alleles we added to the germline collection were all present at greater than 30\% for the IGHV gene in question.

\subsection*{Substitution models, fitting and analysis}
The setting of B cell affinity maturation is substantially different than that typically encountered in molecular evolution studies, and hence there are some differences between our model fitting procedure compared to common practice.
For B cell receptors outside of nontemplated insertions, the root state is the V, D, and J genes encoded in the germline from which a sequenced BCR derives.
Thus, we analyze substitutions that have occurred in evolution away from the germline-encoded segments of observed BCR sequences, ignoring sites comprising nontemplated insertions.
The CDR3 region of an antibody is generally sufficient to uniquely identify its specificity~\citep{Xu2000-gl}.
Although there are certainly some clones in our data set that derive from a single rearrangement event but differ due to somatic hypermutation, the probability that a given pair of distinct sequences derives from a single common ancestor is small: targeted searches for clonally related antibodies during infection have identified them at 0.003\% to 0.5\%\ \citep{Zhu2013-cg}.
It is a substantial challenge to statistically infer which sequences sit together in a clonal lineage and then to perform phylogenetic analysis on such a large data set (see work by \cite{Jiang2013-kj,Kepler2013-sy,Kepler2014-jy}) and performing this analysis incorrectly could bias our results.
Additionally, we encountered significant computational barriers analyzing the volume of sequences available, and adding phylogenetic structure to our analysis may have made the analysis computationally prohibitive even if we had the lineage structure in hand (we believe this is the largest number of sequences from a single data set analyzed in selection study to date).

For these reasons, our analyses were performed on a set of pairwise alignments, each representing a two taxon tree containing an observed sequence and its best scoring germline sequence according to Smith-Waterman alignment.
This is equivalent to using a rooted ``star'' tree where the root state is known.
This assumption allowed us to focus our attention on the selection inference problem.

Substitution models are summarized in Table~\ref{TABLEmodelList} and described in detail here.
We will use $n$ for the number of observed sequences.
Our models are characterized by three components.
First, the subscript of $t$ describes how branch length assignments are allowed to vary across segments of a single sequence.
The $t_i$ model allows branch lengths to vary independently, resulting in $3n$ parameters.
The $t_r$ model has two global per-segment multipliers to define the branch lengths (see, e.g. Fig.~\ref{FIGgtrBl}) with the V segment rate fixed at 1, resulting in $n+2$ parameters.
The subscript of $Q$ describes how rate matrices are fit.
The $Q_i$ model allows an independent global GTR rate matrix for each segment, with a total of 24 parameters.
The $Q_r$ model just has one GTR rate matrix overall, with 8 parameters.
The subscript of $\Gamma$ denotes how across-site substitution rate variation is modeled in terms of a four category discrete gamma distribution~\citep{Yang1994-ik}.
The $\Gamma_i$ model allows an independent rates across sites parameter for each sequence, with 3 parameters.
The $\Gamma_s$ has a global rates across sites parameter, with 1 parameter.
Given these choices concerning how the data was partitioned and parametrized, the standard phylogenetic likelihood function was used as described in the original literature
\citep{Felsenstein1981-zs,Tavare1986-nb,Yang1994-pm} and in books (e.g.\ \citet{Salemi2009-tw,Felsenstein2004-pd}).

Maximum likelihood values of substitution model parameters and branch lengths were estimated using a combination of Bio++ \citep{Gueguen2013-hq} and BEAGLE \citep{Ayres2011-hi}, with model optimization via the BOBYQA algorithm \citep{Powell2009-wy} as implemented in NLopt \citep{johnson2010nlopt}, and branch length optimization via Brent's method \citep{brent1973algorithms}.
Optimization alternated between substitution model parameters and branch lengths until the change in log-likelihood at a given iteration was less than 0.001.
Our software to perform this optimization is available from \url{https://github.com/cmccoy/fit-star}.

For the principal components analysis on substitution matrices, we first obtained the median branch length $\hat{t}$ across all sequences for all individuals.
We then calculated the corresponding transition matrix for each model given equiprobable starting state: $e^{Q \hat{t}} \diag(0.25)$.
These were then projected onto the first two principal components, adapting suggestions for doing PCA in the simplex \citep{Aitchison1983-if}.
Specifically, each row of these matrices, as a discrete probability distribution, is a point in the simplex.
Hence we applied a centered log transformation to each row of this matrix using the \texttt{clr} function of the R package \texttt{compositions} \citep{Van2008-yr}, and followed with standard principal components analysis.

To compare distance between inferred models and sequence distance, we calculated the Hamming distance between all pairs of V genes using the alignment available from the IMGT database \citep{Lefranc2008-kg}.
To obtain distances between models, we calculated the Euclidean distance calculated between pairs of the transformed probability vectors.

\subsection*{Selection analysis}
Because of the large volume of sequences to analyze, we also needed a mechanism to detect selection that could be run on over 15 million sequences, most of which did not share evolutionary history.
Classical means of estimating selection by codon model fitting \citep{Goldman1994-im,Muse1994-yv} could not be used, even in their most recent and much more efficient incarnation \citep{Murrell2013-zf}.
Instead, we used the renaissance counting approach \citep{Lemey2012-ja}, which we modified to work under varying levels of coverage.
A key part of the renaissance counting approach is an empirical Bayes regularization procedure \citep{Robbins1956-cn}.
This procedure uses the entire collection of sites to inform substitution rate estimation at each site individually, effectively sharing data across sites, allowing inference at sites which either display few substitutions or have less sequencing coverage.
We note that obtaining precise per-site selection estimates for hundreds of genes requires a large quantity of sequence data like what we have here: the read coverage decrease on the 5' end of the V gene correspondingly increases the width of the error bar (Figure~\ref{FIGdndsNsExample}, see \cite{figshare-plots}), resulting in a decrease of power for selection regime classification (Figure~\ref{FIGdndsBySite}).

\subsubsection*{Bayesian inference of selection on a star-shaped phylogeny}
To determine the site-specific selection pressure for each V gene, we extended the counting renaissance method, described in \citep{Lemey2012-ja}, to accommodate pairwise analyses of a large number of sequences with a known ancestral sequence and non-constant site coverage.
The counting renaissance method starts by assuming a separate HKY substitution model \citep{Hasegawa1985-sm} for each of the three codon positions and uses Markov chain Monte Carlo (MCMC) to approximate the posterior distribution of model parameters that include substitution rates and phylogenetic tree with branch lengths.
Since in our analyses we assumed that query sequences are related by a star-shaped phylogeny, our model parameters included only HKY model parameters and branch lengths leading to all the query sequences.
Moreover, we fixed the parameters of the HKY model, along with the relative rates between codon positions, to the maximum likelihood estimates produced using the whole dataset.
We note that we could have fit per-codon-position GTR models and used them for stochastic mapping, however such a model would still be substantially misspecified compared to a codon model and thus we decided to follow \citep{Lemey2012-ja} and use HKY for the mapping.
\textit{A priori}, we assumed that  branch lengths leading to the query sequence independently follow an exponential distribution with mean 0.1.
We performed 20,000 iterations of MCMC, scaling the branch length leading to the observed sequence at each iteration, and sampling every 40 iterations to generate a total of 500 samples.
Given each posterior sample of query branch lengths, the counting renaissance method draws a sample of ancestral substitutions conditional on the observed data using a simple per-codon-position nucleotide model; the resulting sampled ancestral sequences are then used to count synonymous and nonsynonymous mutations.

\subsubsection*{Sampling codon substitutions}
For each unique read, for each codon position $l$ and posterior sample $j$, counts of synonymous ($C_{jl}^{(S)}$) and nonsynonymous ($C_{jl}^{(N)}$) substitutions at each site were imputed using stochastic mapping as described above.
\par
For $N$ MCMC iterations based on an alignment of $L$ codons, the result of this procedure was two $N \times L$ matrices, each containing the number of synonymous and nonsynonymous events at each codon position in each posterior sample.
Counts of each substitution type along with the total branch length for each site were aggregated across sequences from the same gene by element-wise addition.
This took about 5 days on an 194 core cluster launched on Amazon Web Services using \texttt{starcluster} (\url{http://star.mit.edu/cluster/}).

\subsubsection*{Empirical Bayes regularization}
The varying length of the CDR3, combined with short observed sequences, leads to quite skewed coverage of sites stratified by gene.
We modified the empirical Bayes regularization procedure of the original counting renaissance method \citep{Lemey2012-ja} to account for varying depth of observation as follows.
First, we define a branch length leading to query sequence $i$ for site $l$ as

\begin{equation*}
t_{il} =
\begin{cases}
  t_{i},& \text{if any residues in the observed sequence $i$}\\
        & \qquad \text{align to codon position } l\\
  0,     & \text{otherwise}
\end{cases}
\end{equation*}

We assume that substitution counts for site $l$ come from a Poisson process with rate $\lambda_l t_l$:

\begin{equation*}
C_l \sim \Poisson(\lambda_l t_l),
\end{equation*}
where $t_l = \sum_{i=1}^n t_{il}$.

As in the original counting renaissance, we assume that the site-specific rates $\lambda_l$ come from a Gamma distribution with shape $\alpha$ and rate $\beta$:

\begin{equation*}
\lambda_l \sim \mathrm{Gamma}(\alpha, \beta).
\end{equation*}

We fix $\alpha$ and $\beta$ to their maximum likelihood estimates $\ahat$ and $\bhat$ by treating sampled branch lengths and counts as fixed and
maximizing the likelihood function
\begin{equation}
  \mathcal{L}(\alpha, \beta) = \left(\frac{\beta^\alpha}{\Gamma(\alpha)}\right)^L \prod_l \frac{t_l^{C_l}}{\Gamma(C_l+1)} \frac{\Gamma(C_l + \alpha)}{(t_l + \beta)^{C_l + \alpha}}.
  \label{eq:marginal}
\end{equation}
We provide a derivation of this likelihood function in the Supplementary Methods.
In contrast to \citep{Lemey2012-ja}, we do not have closed-form solutions for the maximum likelihood or method of moments estimators of $\alpha$ and $\beta$ in this slightly more complex setting.
However it does not add a substantial computational burden to estimate these parameters numerically via the BOBYQA optimizer \citep{Powell2009-wy}.

Given $\ahat$ and $\bhat$, we draw rates $\lambda_l$ from the posterior:
\begin{equation}
  \lambda_l \mid C_l \sim \mathrm{Gamma}(C_l + \ahat, t_l + \bhat),
  \label{eq:regularization}
\end{equation}
derived in the Supplementary Methods.

Estimation of $\alpha$ and $\beta$ by maximizing likelihood (\ref{eq:marginal}) fails when the sample variance of the observed counts $C_1 \ldots C_L$, weighted by the site-specific branch length sums, $t_1 \ldots t_L$,  is less than the corresponding weighted sample mean.
In these cases, we assume that the observed counts are drawn from Poisson distributions with site-specific rate $\lambda t_l$:

\begin{equation*}
  C_l \sim \mathrm{Poisson}(\lambda t_l),
\end{equation*}

where here $\lambda$ is shared across sites, and is estimated from the data by maximizing the likelihood

\begin{equation*}
  L(\lambda) = \prod_l^L \frac{(\lambda t_l)^{C_l}}{C_l!} e^{-\lambda t_l}.
\end{equation*}

\subsubsection*{Simulations}
To validate this method, we simulated 1000 sequences of 100 codon sites each under the GY94 model and a star-like phylogeny with branch lengths fixed to 0.05 using piBUSS \citep{Bielejec2013-ig}.
We varied $\omega$ over the alignment, with 85 sites having $\omega = 0.1$, 5 sites having $\omega = 1$, and 10 sites under positive selection - $\omega = 10$.
We next introduced varying coverage over the alignment: sequences were truncated such that no sequences covered the first 10 codons, only half of the sequences had coverage over the next 40 codons, and all sequences covered the remaining 50 codons (Fig.~\ref{FIGsimulationResults}, bottom panel).
Estimates of $\omega$ were more accurate with higher site coverage (Fig.~\ref{FIGsimulationResults}, top panel).
Of note, as a result of the empirical Bayes regularization, even some sites with no coverage were classified as being under purifying selection.
In all other analyses, we only report $\omega$ estimates for sites covered by at least 100 sequences.
Since the starting state is always the germline amino acid, no classifications can be made for sites which are Tryptophan or Methionine in the germline, as all mutations are nonsynonymous for codons encoding those amino acids.

\subsubsection*{Site-specific estimates of $\omega$}
In \citep{Lemey2012-ja}, the authors arrive at site-specific estimates of $\omega_l$ by comparing data-conditioned ($C$) rates $\lambda_l$ of nonsynonymous ($N$) and synonymous ($S$) substitutions, each normalized by an ``unconditional rate'' ($U$): $\omega_l^{RC} = \frac{\lambda_l^{(N,C)} / \lambda_l^{(N,U)}}{\lambda_l^{(S,C)} / \lambda_l^{(S,U)}}$.
As SHM is highly context-specific, we chose to use rates inferred from out-of-frame rearrangements in place of the unconditional rates, as these more accurately represent the mutation rates in the absence of selection:

\begin{equation}
  \omega_l = \frac{\lambda_l^{(N,P)} / \lambda_l^{(N,O)}}{\lambda_l^{(S,P)} / \lambda_l^{(S,O)}},
  \label{eq:omega}
\end{equation}

where $P$ and $O$ refer to productive and out-of-frame rearrangements, respectively.

\subsubsection*{Implementation}
We used the BEAST \citep{Drummond2012-ek} implementation of the counting renaissance procedure to sample counts for both synonymous and nonsynonymous substitutions at each site.
We extended BEAST version 1.8.0 to generate ``unconditional'' counts using the germline state as the starting state for simulating along the edge to the query as described above. 
This process (sampling substitutions for each sequence, then combining counts from sequences mapping to the same IGHV) provides a natural setting for parallelization via the map-reduce model of computation; we used the Apache Spark \citep{Zaharia2010-xr} framework to distribute work across a cluster running on Amazon EC2.
Our software to perform this analysis is available from \url{https://github.com/cmccoy/startreerenaissance}.

\subsection*{Structural analysis}
For each of the eleven most frequently occurring V genes, we identified the closest structure in the Protein Data Bank (PDB)~\citep{Berman2000-pw} using BLAST~\citep{Altschul1997-tc}.
Structures were visualized using PyMOL~\citep{Delano2002-rr}.
We calculated the normalized solvent-accessible surface area (SASA) for each amino acid position using ROSETTA3~\citep{Leaver-Fay2011-my} and normalized these values by dividing them by the fully exposed SASA of the given residue type in an extended chain.
Wilcoxon rank-sum tests~\citep{hollander1973nonparametric} between all pairs of selection classifications (negative, neutral, positive) were used to assess whether the normalized SASA differed between groups\@.
p-values were Bonferroni-corrected~\citep{Holm1979-tn} to account for multiple comparisons.

The details of our computational methods are available in the \emph{Supplementary Methods}.

\section*{Data Accessibility}
Data is available at \url{http://adaptivebiotech.com/link/mat2015}.
We have made an Amazon Machine Image (AMI)~\citep{amazon2014-faq} available pre-loaded with our analysis pipeline and some example data.
To use it, launch an instance of AMI \texttt{ami-ab295b9b} in the \texttt{us-west-2} region and log in as user \texttt{ubuntu} (no password needed: authentication by public key).

\section*{Funding}
This work was supported by the National Institute of Health [R01-AI107034 to V.N.M., R01-GM11324601 to F.A.M., R01-AI103981 to Dr. Julie Overbaugh] and the National Science Foundation [DMS-0856099 to V.N.M]; C.O.M. and F.A.M. supported in part by the University of Washington eScience Institute through its Seed Grants program in Translational Health Sciences, and by a 2013 new investigator award from the University of Washington Center for AIDS Research (CFAR), an NIH funded program under award number P30AI027757 which is supported by the following NIH Institutes and Centers: NIAID, NCI, NIMH, NIDA, NICHD, NHLBI, NIA, NIGMS, and NIDDK\@.

\section*{Authors' contributions}
CM designed and carried out the computational analysis, developed the statistical methods, and wrote the paper;
TB designed the computational analysis, developed the statistical methods, and wrote the paper;
VM designed the computational analysis, developed the statistical methods, and wrote the paper;
PB performed structural analysis;
HR helped define questions and provided the data;
FM designed the computational analysis, developed the statistical methods, and wrote the paper.

\section*{Competing interests}
HR owns stock in and consults for Adaptive Biotechnologies.
The other authors have no competing interests.

\section*{Acknowledgements}
The molecular work for this project was done by Paul Lindau in the laboratory of Phil Greenberg, and was supported by a grant from the W. M. Keck Foundation.
The authors are grateful to Bryan Howie and William DeWitt III for helpful discussions.

\clearpage

\notarxiv{
\section*{Table and Figure Legends}
\TABLEmodel
\FIGgtrBl
\FIGcompPca
\FIGdndsNsExample
\FIGdndsBySite
\FIGdndsDistribution
\FIGstructure
\FIGoverview
\beginsupplement

\clearpage

\section*{Supplementary Table and Figure Legends}
\TABLEreadCounts
\TABLEnumGermline
\TABLEbaseFrequencies
\FIGgtr
\FIGgtrConsistency
\FIGmodelVsHammingDist
\FIGaa
\FIGdndsConsistency
\FIGsimulationResults
\FIGsasa

\clearpage
}

\bibliographystyle{vancouver}
\bibliography{mebcell}

\notarxiv{\end{document}}

\clearpage

\beginsupplement

\begin{center}

\noindent{\Large Supplementary Material: Quantifying evolutionary constraints on B cell affinity maturation}

\qquad

\noindent {\normalsize \sc
Connor O. McCoy, Trevor Bedford, Vladimir N. Minin, Philip Bradley, Harlan Robins, and Frederick A. Matsen IV}\\
\noindent {
\textit{Philosophical Transactions of the Royal Society B}; doi:10.1098/rstb.2014-0244
}\\
\end{center}

\bigskip

\section*{Supplementary Methods}

\subsection*{Derivation of the Gamma-Poisson marginal likelihood with varying observation depth}
We will use the same notation as in the Materials and Methods section.
Our first task is to write down a likelihood of $\alpha$ and $\beta$ given a collection of counts.
To do so we will marginalize out the rates $\lambda_l$ when they are drawn from a $\mathrm{Gamma}(\alpha, \beta)$ as in the main text.

The likelihood for a single site is (omitting $l$ for now):

\begin{eqnarray*}
  P(C \vert  t, \alpha, \beta) &=& \int_0^{\infty} P(C\vert t, \lambda)P(\lambda\vert \alpha, \beta)d\lambda\\
  &=& \int_0^\infty \frac{(\lambda t)^C e^{-\lambda t}}{C!} P(\lambda\vert \alpha, \beta)d\lambda\\
  &=& \int_0^\infty \frac{(\lambda t)^C e^{-\lambda t}}{C!} \left[ \frac{\beta^\alpha}{\Gamma(\alpha)}\lambda^{\alpha-1} e^{-\beta\lambda} \right] d\lambda\\
  &=& \frac{\beta^\alpha t^C}{C! \, \Gamma(\alpha)} \int_0^\infty \lambda^{C + \alpha - 1} e^{-\lambda(t + \beta)} d\lambda.
\end{eqnarray*}

Letting $\alpha' = C + \alpha$ and $\bprim = t + \beta$, introduce a
normalizing constant for the distribution
$\mathrm{Gamma}(\alpha', \bprim)$:

\begin{eqnarray*}
P(C \vert  t, \alpha, \beta) &=& \frac{\beta^\alpha t^C}{C! \, \Gamma(\alpha)} \frac{\Gamma(\alpha')}{\bprim^{\alpha'}}  \int_0^\infty \frac{\bprim^{\alpha'}}{\Gamma(\alpha')} \lambda^{\alpha' - 1} e^{-\lambda(\bprim)} d\lambda\\
&=& \frac{\beta^\alpha t^C}{C! \, \Gamma(\alpha)} \frac{\Gamma(\alpha')}{\bprim^{\alpha'}}  \int_0^\infty \DGamma(\lambda; \alpha', \bprim) d\lambda.
\end{eqnarray*}

The integral over the support of the Gamma distribution is $1$, so:
\begin{eqnarray*}
  P(C \vert  t, \alpha, \beta) & = & \frac{\beta^\alpha t^C}{C! \, \Gamma(\alpha)} \frac{\Gamma(\alpha')}{\bprim^{\alpha'}} \\
  &=& \frac{\beta^\alpha t^C}{C! \, \Gamma(\alpha)} \frac{\Gamma(C + \alpha)}{(t + \beta)^{C + \alpha}}.
\end{eqnarray*}

The overall marginal likelihood is the product over such sites:
\begin{eqnarray*}
  \mathcal L &=& P(C_1, \ldots, C_L \vert  t_1, \ldots, t_L, \alpha, \beta) =  \prod_l \frac{\beta^\alpha t_l^{C_l}}{C_l! \, \Gamma(\alpha)} \frac{\Gamma(C_l + \alpha)}{(t_l + \beta)^{C_l + \alpha}} \\
  &=& \left(\frac{\beta^\alpha}{\Gamma(\alpha)}\right)^L \prod_l \frac{t_l^{C_l}}{C_l!} \frac{\Gamma(C_l + \alpha)}{(t_l + \beta)^{C_l + \alpha}} \\
  &=& \left(\frac{\beta^\alpha}{\Gamma(\alpha)}\right)^L \prod_l \frac{t_l^{C_l}}{\Gamma(C_l+1)} \frac{\Gamma(C_l + \alpha)}{(t_l + \beta)^{C_l + \alpha}},
\end{eqnarray*}
giving \eqref{eq:marginal}.

\subsubsection*{Posterior for $\lambda$}
Our eventual goal is a regularized posterior estimate of the rates $\lambda_l$.
For a single site, once again dropping $l$:

\begin{equation*}
P(\lambda \vert  C, t, \ahat, \bhat) \propto P(C \vert  \lambda, t) P(\lambda\vert \ahat, \bhat).
\end{equation*}

Substituting in the PDFs for the distributions employed for $C$ and
$\lambda$:

\begin{equation*}
P(\lambda \vert  C, t, \ahat, \bhat) \propto \frac{\bhat^{\ahat} t^C}{C! \, \Gamma(\hat \alpha)} \lambda^{C + \hat \alpha - 1} e^{-\lambda(t + \hat \beta)}.
\end{equation*}

As in the main text, we let $\hat \alpha' = C + \hat \alpha$ and $\hat \bprim = t + \hat \beta$.

\begin{eqnarray*}
P(\lambda \vert  C, t, \ahat, \bhat) &\propto& \frac{\bhat^{\ahat} t^C}{C! \, \Gamma(\ahat)} \frac{\Gamma(\ahat')}{\bhat'^{\ahat'}}  \left[ \frac{\bhat'^{\ahat'}}{\Gamma(\ahat')} \lambda^{\ahat' - 1} e^{-\lambda(\bhat')}\right] \\
  &\propto& \frac{\bhat^{\ahat} t^C}{C! \, \Gamma(\ahat)} \frac{\Gamma(\ahat')}{\bhat'^{\ahat'}} \DGamma(\lambda; \ahat', \bhat') \\
  &\propto& \DGamma(\lambda; \ahat', \bhat'),
\end{eqnarray*}
hence these two probability densities are equal, justifying \eqref{eq:regularization}.

\clearpage

\section*{Supplementary Figures and Tables}

\TABLEreadCounts
\TABLEnumGermline
\TABLEbaseFrequencies
\FIGgtr
\FIGgtrConsistency
\FIGmodelVsHammingDist
\FIGaa
\FIGdndsConsistency
\FIGsimulationResults
\FIGsasa

\end{document}